\documentclass[useAMS,usegraphicx,usenatbib]{mn2e}
\usepackage{times}

\title[MGC: bimodality in the colour-concentration plane]{The Millennium Galaxy
Catalogue: morphological classification and bimodality in the
colour-concentration plane.}

\author[S.P.~Driver et al.]{S.P.~Driver,$^{1}$\thanks{E-mail:
spd@mso.anu.edu.au}
P.~D.~Allen$^{1}$,
Alister.~W.~Graham$^{1}$,
E.~Cameron$^{1}$,
J.~Liske$^2$,
S.C.~Ellis$^3$, \newauthor
N.J.G.~Cross$^4$,
R.~De~Propris$^{5}$,
S.~Phillipps$^{5}$,
W.~J.~Couch$^6$ \\ \\
$^1$Research School of Astronomy and
Astrophysics, Australian National University, Cotter Road, Weston, ACT
2611, Australia \\ 
$^2$European Southern Observatory,
Karl-Schwarzschild-Str.~2, 85748 Garching, Germany \\ 
$^3$Anglo-Australian Observatory, P.O. Box 296, Epping, NSW 2121, Australia \\
$^4$ Institute for Astronomy, University of Edinburgh, Royal Observatory, Edinburgh, EH9 3HJ, Scotland \\
$^5$Department of Physics, University of Bristol,
Tyndall Avenue, Bristol, BS8 1TL, UK \\
$^6$School of Physics, University of New South Wales, Sydney, NSW 2052, Australia}

\newcommand{\mpas}{~mag~arcsec$^{-2}$}

\newcommand{\bmgc}{B_{\mbox{\tiny \sc MGC}}}

\begin{document}

\date{Accepted
...... Received .....}

\pagerange{\pageref{firstpage}--\pageref{lastpage}} \pubyear{2004}

\maketitle

\label{firstpage}

\begin{abstract}
Using 10 095 galaxies ($B < 20$ mag) from the Millennium Galaxy
Catalogue, we derive $B$-band luminosity distributions and selected
bivariate brightness distributions for the galaxy population
subdivided by: eyeball morphology; S\'ersic index ($n$); 2dFGRS
$\eta$-parameter; rest-$(u-r)$ colour (global and core); MGC continuum
shape; half-light radius; (extrapolated) central surface brightness;
and inferred stellar mass-to-light ratio. All subdivisions extract
highly correlated sub-sets of the galaxy population which consistently
point towards two overlapping distributions: an old, red, inert,
predominantly luminous, high central-surface brightness subset; and a
young, blue, star-forming, intermediate surface brightness
subset. {\it A clear bimodality in the observed distribution is seen
in both the rest-$(u-r)$ colour and $\log(n$) distributions.} Whilst
the former bimodality was well established from SDSS data
\citep{strateva01}, we show here that the rest-$(u-r)$ colour
bimodality becomes more pronounced when using the core colour as
opposed to global colour. The two populations are extremely well
separated in the colour-log($n$) plane. Using our sample of 3 314 ($B
< 19$ mag) eyeball classified galaxies, we show that the
bulge-dominated, early-type galaxies populate one peak and the
bulge-less, late-type galaxies occupy the second. The early- and
mid-type spirals sprawl across and between the peaks. This constitutes
extremely strong evidence that the fundamental way to divide the
luminous galaxy population ($M_{\bmgc} -5 \log h < -16$ mag, i.e.,
dwarfs not included) is into bulges (old red, inert, high
concentration) and discs (young, blue, star-forming, low
concentration) and that the galaxy bimodality reflects the two
component nature of galaxies and not two distinct galaxy classes. We
argue that these two-components require two independent formation
mechanisms/processes and advocate early bulge formation through
initial collapse and ongoing disc formation through splashback, infall
and merging/accretion. We calculate the $B$-band luminosity-densities
and stellar-mass densities within each subdivision and estimate that
the $z \approx 0$ stellar mass content in spheroids, bulges and discs
is $35 \pm 2$ per cent, $18 \pm 7$ and $47 \pm 7$ per cent
respectively.
\end{abstract}


\section{Introduction}
Galaxies exhibit remarkable diversity, and as such it is not clear how
useful collective studies are, {\it i.e.,} those that bunch all
galaxies into a single sample. For instance knowledge of the variation
of the median galaxy size with redshift has no unique interpretation
if multiple evolutionary mechanisms exist. Given the established
diversity in luminosity, size, shape, colour, star-formation rate,
metallicity, gas and dust content, it seems highly likely that
multiple evolutionary paths, mechanisms and time-scales do exist. From
a theoretical perspective there are several proposed modes of
evolution: monolithic collapse (\citealp{elbs}; \citealp{sandage90}),
hierarchical merging (\citealp{white}; \citealp{fall}), gas infall
\citep{infall}, satellite accretion \citep{Searle}, secular evolution
(see review by \citealp{kk}) and splashback (\citealp{fukugita}).  The
number of secondary (and mainly environmentally dependent) processes
is even higher, e.g., tidal formation (\citealp{barnes92}),
ram-pressure stripping (\citealp{gunn72}), strangulation
(\citealp{balogh}), harassment (\citealp{moore}), squelching
(\citealp{squelch}), threshing (\citealp{bekki}), and cannibalism
(\citealp{cannibalism}), for example. Each may apply, in differing
degrees at different times with environmental and initial condition
dependencies. A key question is whether any of the readily observable
properties of the galaxy population today contains a clear imprint of
these distinct formation mechanisms. One purely empirical way to
attempt to identify such connections is to look for naturally
occurring sub-groupings within some region of galaxy parameter space.

At this point the natural subdivision(s) of galaxies is not entirely
clear (see for example \citealp{blanton}) and the quest to determine a
fundamental basis is ongoing and constitutes a continuation of the
same questions asked by \cite{hubble36rn} and
\cite{zwicky57}. Traditionally the favoured mechanisms have been
eyeball classification against a set of visual criteria
(\citealp{hubble}), resulting in the [Hubble] tuning fork
(\citealp{jeans29}, \citealp{hubble36rn}; \citealp{sandage61}), or
similar schemes (e.g., \citealp{devauc56}; \citealp{vdb76}), and via
light-profile fitting (\citealp{devauc48}, 1959; \citealp{sersic},
1968; \citealp{ken}).  Both systems remain in common usage and are now
being routinely applied to large datasets in an automated manner. For
example, Artificial Neural Networks are now used to implement eyeball
classification (\citealp{avi}; \citealp{odewahn}; \citealp{ball04})
and public codes such as GIM2D (\citealp{gim2d}) and GALFIT
(\citealp{galfit}) are now available for automated bulge-disc
decomposition or single S\'ersic profile fitting.  Much of the drive
to pursue these methods now comes from our ability to apply coarse
structural measurements at any redshift via high-resolution
space-based observatories (see for example \citealp{driver95a}, 1995b;
\citealp{lilly98}; \citealp{goods}).

Significant effort is also being invested in exploring alternatives
such as: the joint luminosity-surface brightness(size) plane (LSP; see
\citealp{driver05} and references therein), the colour-luminosity
plane (CLP; \citealp{strateva01}; \citealp{baldry};
\citealp{balogh04a}; \citealp{faber}) --- both of which have
traditional roots\footnote{ \cite{hubble} showed a relation between
luminosity and diameter, (see also \citealp{binggeli}), while
\cite{holmberg58} and \cite{sv78} showed a colour magnitude relation
for ellipticals.} --- spectral classification (\citealp{madgwick}),
line strengths (\citealp{lin96}; \citealp{kauffmann}, \citealp{balogh04b}),
Concentration/Asymmetry/Smoothness parameters (\citealp{conselice}),
the Gini coefficient (\citealp{abrahams}, \citealp{lotz}), Principle
Component Analysis (\citealp{ellis}), Shapelet analysis
\citep{shapelet}, and Fourier decomposition (\citealp{odewahn3}) for
example. At present no single approach stands above the rest and all
are open to various criticisms. For example eyeball morphologies are
subjective, the LSP and CLP are too coarse,
Concentration/Asymmetry/Smoothness is susceptible to short term
transitory effects (minor mergers, interactions, episodic star-burst
etc), the Gini coefficient and PCA analysis are unlikely to lead to a
straightforward connection to the underlying physics, not all galaxies
are readily profiled by S\'ersic or S\'ersic plus exponential models,
and Fourier decomposition requires high signal-to-noise and is hence
not applicable beyond the local universe. In addition no method has
been shown to have a deep physics basis, although galaxy profiling
probably comes closest (see \citealp{king63} and \citealp{fall} for
example).

Moving beyond classification issues, the standard method for
representing the collective galaxy population has typically been via
the luminosity distribution (e.g., \citealp{hubble36}; Binggeli,
Sandage \& Tammann 1988) and its adopted analytical representation,
the Schechter luminosity function (\citealp{schechter}). Contemporary
measurements of the global $~B$-band galaxy luminosity distribution
started with the CfA slice \citep{cfa} followed by the analysis of
\cite{eep} and have culminated in recent results from the ESO Slice
project (\citealp{esp}), the Two-degree Field Galaxy Redshift Survey
(\citealp{norberg02}), the Sloan Digital Sky Survey
(\citealp{blanton03}), and the Millennium Galaxy Catalogue
(\citealp{driver05}) and references therein. Generally these surveys
now concur on the space-density of luminous galaxies ($M_{\bmgc} -5
\log h < -16$ mag), with the space-density of dwarf systems
($M_{\bmgc} -5 \log h > -16$ mag) essentially unconstrained
(\citealp{tully}, although see \citealp{blanton05} for a considered
attempt based on SDSS data). Data from each of these surveys has also
been used to provide galaxy luminosity distributions subdivided by
various criteria, e.g., spectral type \citep{madgwick}, colour
\citep{esp}, galaxy light-profile shape \citep{blanton03} and
morphology \citep{nakamura}. These and earlier attempts to subdivide
the galaxy population have been comprehensively summarised by
\cite{lapparent}, who highlights the complexity and confusion that can
arise through the comparison of galaxy LFs subdivided by differing
criteria (i.e., although morphology, spectral type, colour and profile
shape have long been known to be correlated, the terms are not
synonymous). This once again leads back to the question as to what is
the fundamental way to divide the galaxy population if at all?  This
is not an easy question to answer but empirically one can address it
by highlighting the division that leads to maximum variance in the
recovered Schechter function parameters as well as by highlighting
{\it modality} in joint distributions. Recent studies have identified
bimodalities in the colour-magnitude plane (\citealp{strateva01},
\citealp{ball}), the stellar age-stellar mass plane
(\citealp{kauffmann}), the luminosity-surface brightness plane
(\citealp{driver05}) and the luminosity-size plane (\citealp{shen03},
\citealp{blanton}). While the nature of these bimodalities is no doubt
of common origin it remains unclear as to which plane is more
fundamental and what these bimodalities might be telling us. Our
argument, which we develop throughout this paper, is that they all
reflect the {\it two component} nature of galaxies (as opposed to two
distinct classes of galaxies) and that the fundamental division is
between the disc and bulge components, each of which may have a
distinct formation mechanism acting over two distinct eras.

Using the Millennium Galaxy Catalogue (MGC; \citealp{mgc1}) we explore
the coarse global properties of luminous galaxies ($M_{\bmgc} -5 \log
h < -16$ mag). In Section 2 we summarize the MGC and describe the
morphological classification process which supplements the colour,
spectral and structural parameters which are described in detail
elsewhere (see \citealp{driver05}, \citealp{allen05}). In Section 3 we
explore the distribution of galaxy morphological type, MGC continuum type,
rest-$(u-r)$-colour, global S\'ersic profile shape and derive the
luminosity distributions divided along natural boundaries. In Section
4 we continue the exploration by examining selected bivariate
distributions. Implications for galaxy formation are discussed in
Section 5. We use $\Omega_{M}=0.3, \Omega_{\Lambda}=0.7$, $H_{0}=100
h$ km s$^{-1}$ Mpc$^{-1}$ throughout.

\section{The Millennium Galaxy Catalogue}
\begin{figure*}
\centering\includegraphics[width=\textwidth]{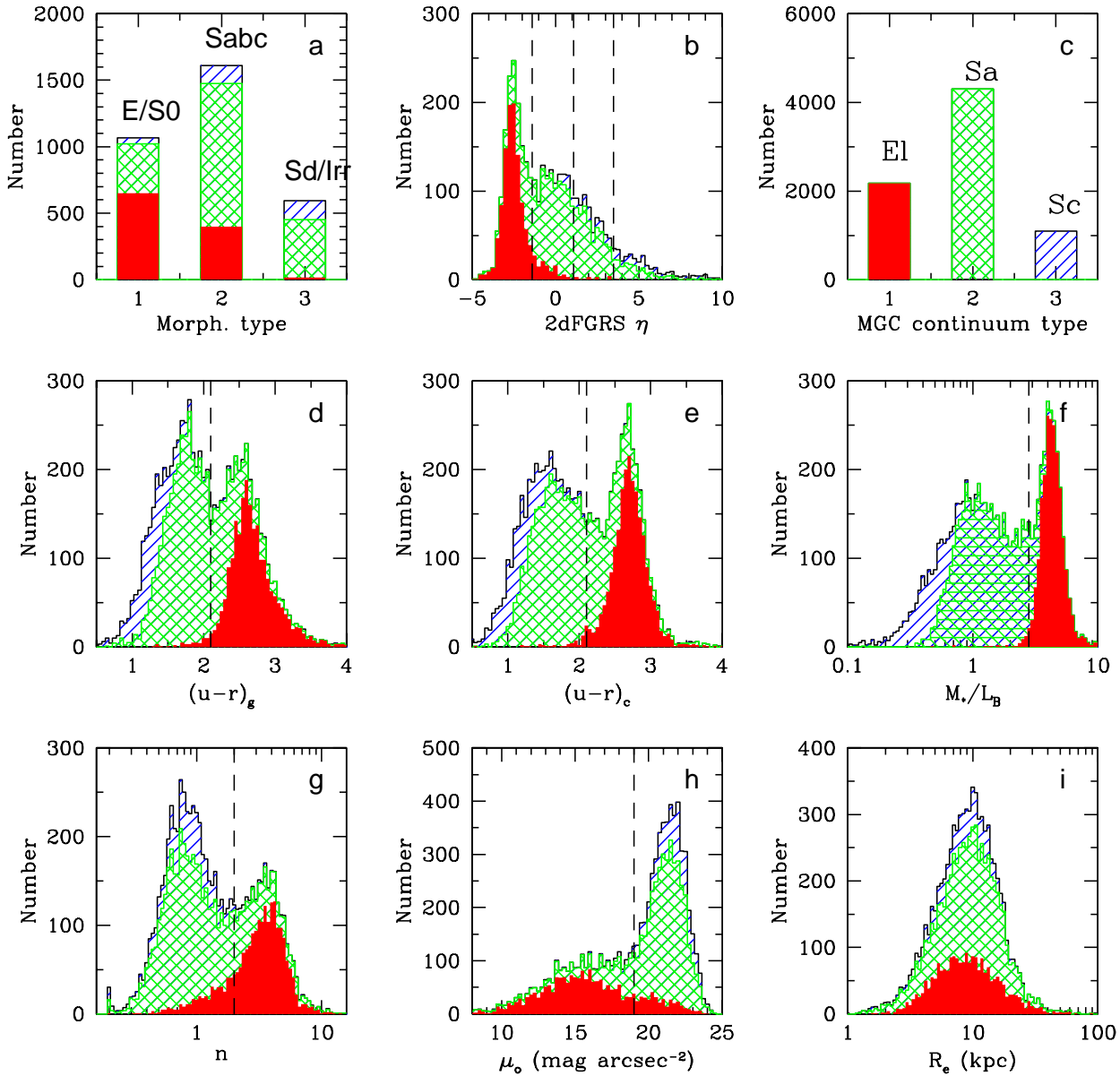}
\caption{The various criteria by which the MGC can be subdivided (as
indicated, see also Sections 2.1 -- 2.8 ). The shading represents the
location of MGC galaxies classified based on their continuum shape as either
El (solid), Sa (hashed) or Sc (diagonal lines).}
\label{panel2}
\end{figure*}

\begin{figure*}
\centering\includegraphics[width=\textwidth]{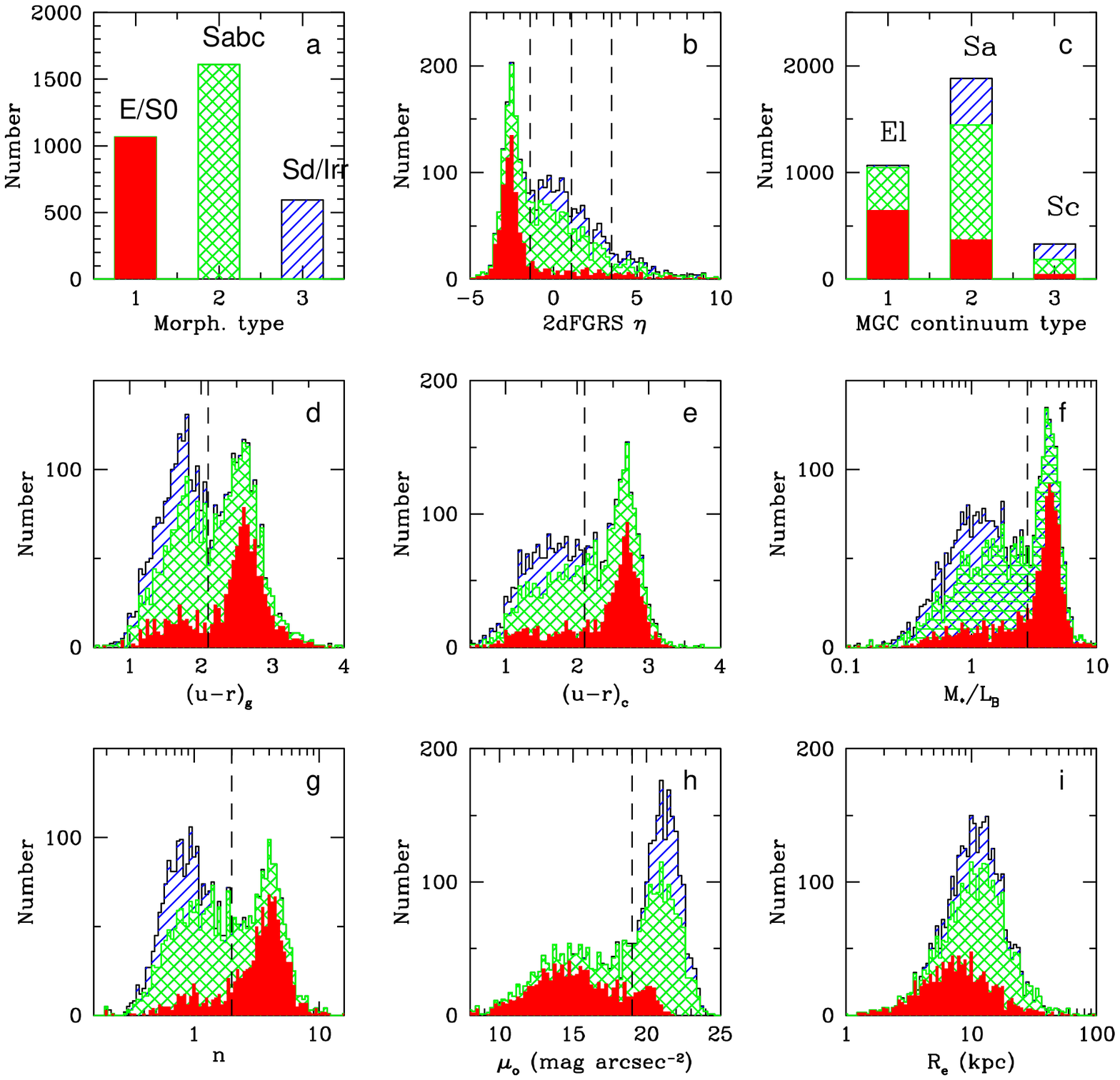}
\caption{As for Fig.~\ref{panel2} except for galaxies with $\bmgc <
19$ mag only throughout. The shading now highlights how the eyebal
morphological classifications into either E/S0s (solid), Sabcs
(hashed) or Sd/Irrs (diagonal lines).}
\label{panel1}
\end{figure*}

The Millennium Galaxy Catalogue (MGC; \citealp{mgc1}) is a deep
($\mu_{\mbox{\tiny \sc limit}} = 26.0~\bmgc$ \mpas) survey of a 37.5
deg$^2$ region of sky, $\sim0.6$ deg wide and extending from
$\sim10^h$ to $\sim14^h50^m$ along the J2000.0 equator. The imaging
survey was conducted with the Wide Field Camera installed on the 2.5m
Isaac Newton Telescope in La Palma and the survey's design, execution,
reduction, object detection, and preliminary analysis are described in
\cite{mgc1}.  The MGC lies within the Two Degree Field Galaxy Redshift
Survey (2dFGRS) Northern Galactic Cap region and the Sloan Digital Sky
(SDSS) Early Data Release Region. A detailed comparison of the MGC
with the much larger but shallower SDSS (Data Release One, DR1,
\citealp{abazajian03}) is described in \cite{mgc3}. The spectroscopic
extension (MGCz), which built upon the redshifts provided by the 2dFGRS
and SDSS, is described in \cite{driver05}. The {\sc mgc-bright}
imaging catalogue contains 10 095 galaxies ($13.0 < \bmgc < 20.0$
mag), for which redshifts have now been obtained for 9 696 resulting
in a global completeness of 96 per cent.  In this paper we
occasionally restrict ourselves to galaxies brighter than $\bmgc <
19.0$ mag. In these cases the sample size is reduced to 3 492 galaxies
for which redshifts have been obtained for 3 487 indicating a global
completeness of 99.9 per cent. Selection biases within the catalogue
and their treatment are extensively discussed in \cite{driver05}. The
MGC photometric system ($\bmgc$) is calibrated to Vega and common
filter conversions are shown in \cite{mgc1} and Appendix A of
\cite{mgc3}. The MGC catalogue used in this paper is available
online\footnote{http://www.eso.org/$\sim$jliske/mgc/} and includes 120
parameters per galaxy derived from our own software, GIM2D and through
matching to the 2dFGRS and SDSS-DR1 datasets. We now describe the
parameters used in this paper and their derivations starting with
eyeball morphology. The resulting observed distributions (i.e.,
uncorrected for volume bias) of each parameter (morphology, 2dFGRS
$\eta$, MGC continuum classification, global and core colours,
mass-to-light ratio, S\'ersic index, central surface brightness and
half-light radius), are shown on Figs.~\ref{panel2}~and~\ref{panel1} for
the $\bmgc < 20, M_{\bmgc} - 5\log h < -16$ mag and $\bmgc < 19,
M_{\bmgc} - 5\log h < -16$ mag samples respectively.

\subsection{The MGC morphologies}
We have derived the morphological classifications for the 3 314
galaxies that lie within the MGC size and surface brightness selection
boundaries (see \citealp{driver05}) with $\bmgc < 19$ mag through
eyeball classification (SPD). This is a subjective process and was
achieved by visual inspection of batches of 20 grey scale images
displayed over three surface brightness ranges: 19.0 -- 26 \mpas,
20.5 -- 26 \mpas, 22.0 -- 26 \mpas. The three visual
classes are defined as follows:

~

\noindent
$\bullet$ E/S0: Smooth, highly concentrated symmetrical systems with
no obvious spiral-arm disc component.

~

\noindent
$\bullet$ Sabc: Clear two-component bulge-disc system, generally
smooth and symmetrical.

~

\noindent
$\bullet$ Sd/Irr/Pec: Disc only (i.e., no obvious single core),
asymmetric, highly disturbed or multiple-core systems.

~

\begin{figure*}
\centering\includegraphics[width=16.0cm]{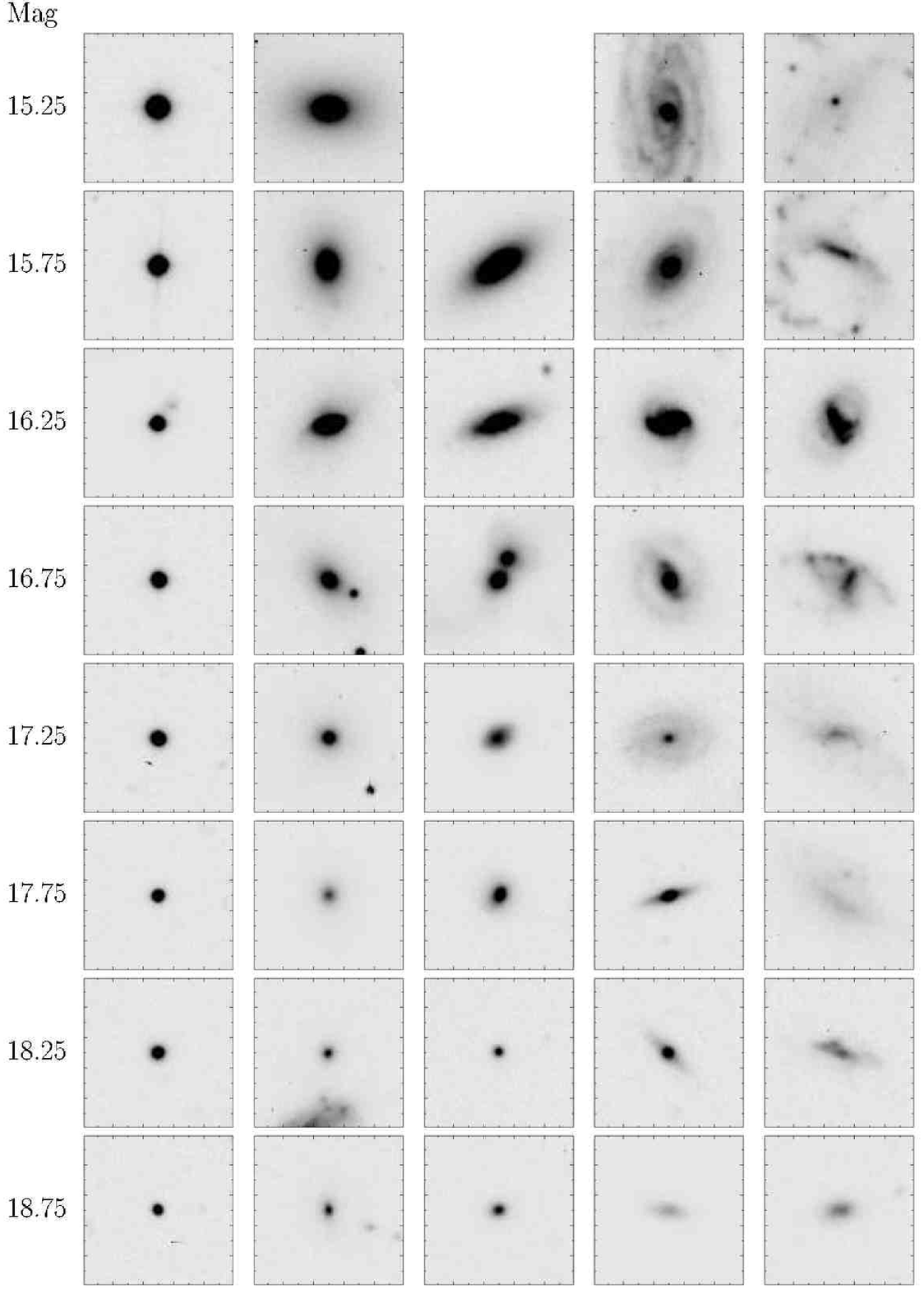}
\caption{A random sample of our postage stamp images
for representative examples of (left to right) stars, red ellipticals, blue
ellipticals (dEs?), spirals, and late-type irregulars at various
magnitude intervals (as indicated). Faintwards of $\bmgc=19$ mag
morphological classification becomes intractable and higher-resolution
data in better seeing is required.}
\label{platemag}
\end{figure*}

\noindent
These classifications are consistent with our deeper {\it Hubble Space
Telescope} studies of more distant galaxies (\citealp{driver95a},
1995b, 1998, 2003; \citealp{driver99}; \citealp{odewahn2};
\citealp{cohen}), which are based on comparable physical resolution
and signal-to-noise data. The entire classification process was
undertaken by SPD over a period of some weeks and repeated several
times until no reclassifications were necessary. Fig.~\ref{platemag}
shows representative examples of each class including stars (see
\citealp{mgc1} for details of the star-galaxy separation). Note that
images of both red and blue 'ellipticals' are shown and these are
discussed in Section 3.2.  Table~\ref{ncounts} contains the
morphological number counts which can be compared to our earlier and
deeper {\it HST} studies listed above.

\begin{table*}
\caption{Morphological number-counts from the MGC.}
\label{ncounts}
\begin{tabular}{ccccc} \hline \hline
$\bmgc$ (mag) & log N[All] & log N[E/S0] & log N[Sabc] & log N[Sd/Irr] \\
                  & & & & \\ \hline
13.75& -1.012& -1.188& -1.489& --- \\
14.25& -0.790& -1.489& -1.012& -1.489 \\
14.75& -0.489& -1.012& -0.644& ---    \\
15.25& -0.343& -0.790& -0.586& -1.489 \\
15.75&  0.123& -0.535& -0.074& -0.711 \\
16.25&  0.396& -0.127& 0.200& -0.790 \\
16.75&  0.677&  0.226&  0.373& -0.147 \\
17.25&  0.894&  0.385&  0.640&  0.015 \\
17.75&  1.228&  0.717&   0.911&  0.547 \\
18.25&  1.474&  0.978&  1.143&  0.802 \\
18.75&  1.686&  1.213&  1.357&  0.974 \\
19.25&  1.901&  --- &  ---  & ---   \\
19.75&  2.122&  --- &  ---  & --- \\ \hline
\end{tabular}
\end{table*}

\subsection{The 2dFGRS $\eta$ parameter}
The 2dFGRS $\eta$ parameter is defined in \cite{madgwick} and is a
linear combination of PC1 and PC2 from a Principle Component Analysis
of the spectra. PC1 contains information on the emission and
absorption line strengths and the continuum (with roughly equal
weights) and PC2 contains information on just the line strengths. The
linear combination maximises line features and hence provides a
good indicator of current and previous star-formation. The reader is
referred to fig.~7 of \cite{madgwick} for canonical examples of
$\eta$ type 1, 2, 3, \& 4 spectra.

\subsection{MGC continuum classification}
\begin{figure}
\centering\includegraphics[width=\columnwidth]{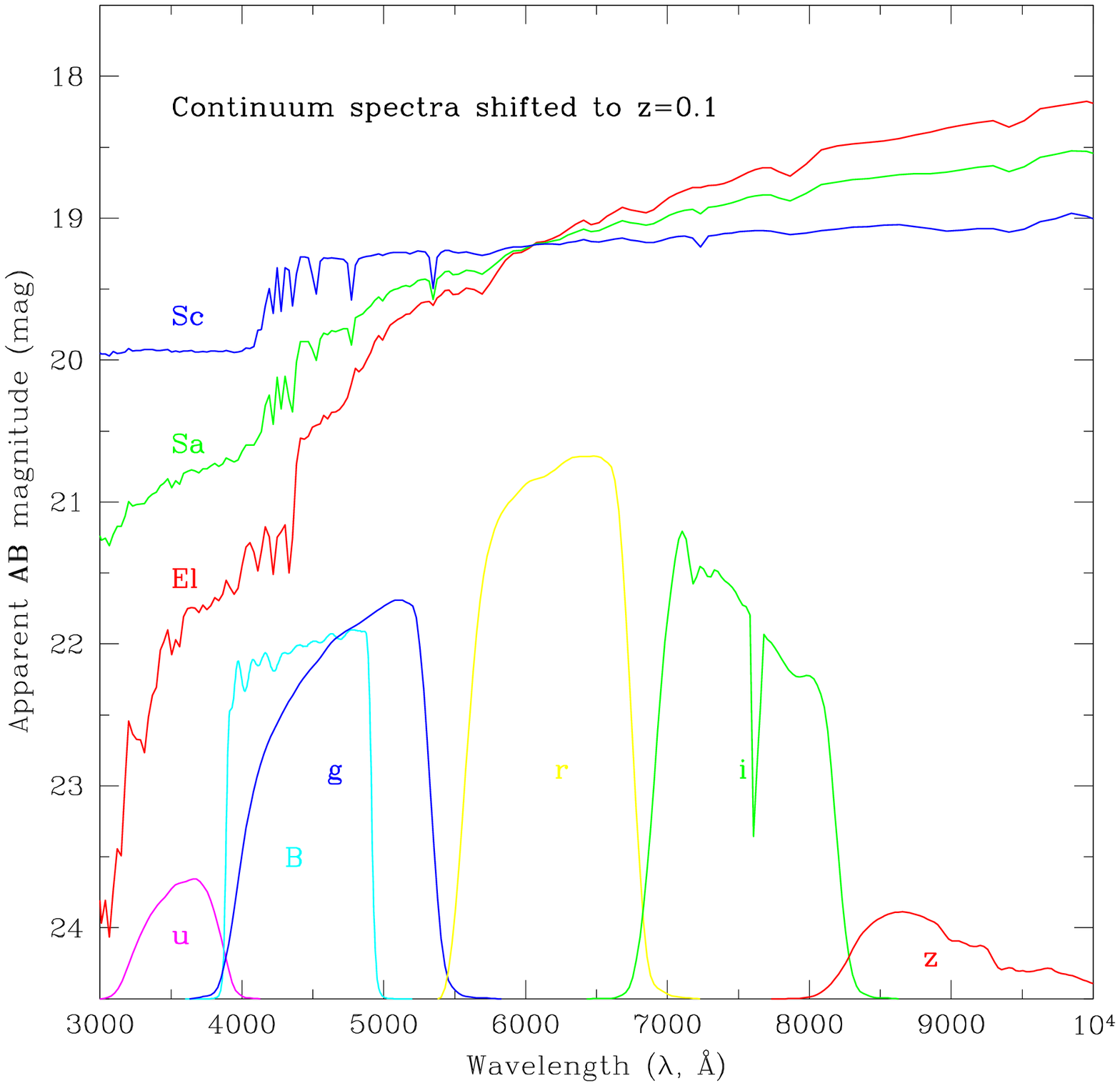}
\caption{The three adopted continuum spectra used in our fitting
process. The spectra are shifted to a redshift of 0.1 and are
arbitrarily scaled. We also show the transmission curves of the
filters for which photometry is available.}
\label{spectraltypes}
\end{figure}

We divide the sample into three MGC continuum types using our spectral
template fitting code (see \citealp{driver05}) and restricting it to
fit either a 15 Gyr elliptical (El), a 7.4 Gyr early-type spiral (Sa)
or a 2.2 Gyr late-type spiral (Sc). All three continuum spectra are
taken from \cite{poggianti97}. Fig.~\ref{spectraltypes} shows both the
location of the broad band filters and our three canonical continuum
shape spectra shifted to $z = 0.1$.  The optimal fitting spectrum for
each galaxy is obtained as described in \cite{driver05}. In brief, the
flux for the $uBgriz$ filter set is derived for each template at the
known redshift of the object and $\chi^2$-minimisation used to
determine the best fitting template (where the normalisation is
marginalised over). Where no satisfactory fit is found the filters are
dropped in the order of $\bmgc$ (non-SDSS), then $u$ (red leak) and
then $z$ (fringing). The optimal spectra provide an indication of the
continuum shape. In essence the MGC continuum typing is simply a
sophisticated colour cut whioch uses all available colours hence
correlation between continuum type and colours is to be expected.

\subsection{SDSS global and core colours}
Overlap with the SDSS-DR1 enables complimentary $ugriz$ photometry for
99\% of the MGC galaxy sample (see \citealp{mgc3}). To convert these
colours to rest-wavelengths we use the appropriate spectral template
and a simple evolutionary recipe as described in \cite{driver05}. The
SDSS-DR1 colour bimodality was first reported by \cite{strateva01} and
studied in more detail by \cite{baldry}. Its environmental dependency
was explored further in \cite{balogh04a}. These studies typically
adopted the {\it global} $(u-r)$ rest colour.  Here we show both the
{\it global} and the {\it core} $(u-r)$ rest colours (hereafter
$(u-r)_g$ and $(u-r)_c$ respectively). The former is derived from
Petrosian magnitudes and the latter from PSF magnitudes.  One may
argue whether the core-colour is better defined by the PSF magnitude,
by a minimal fixed aperture colour, or even fibre magniutdes which are
degraded to a consistent poor seeing value. Clearly, PSF magnitudes
are more appropriate for ultra-compact sources, and a fixed aperture
slightly larger than the seeing more appropriate for extended
sources. However as the cores are often barely resolved we essentially
opt for the smallest credible measurement which is the PSF
magnitude. While this is somewhat profile dependent an equivalently
small fixed aperture would be biased by the wavelength dependency of
the PSF size.  Both of the colour distributions are seen to be
strongly bimodal (see Fig.~\ref{panel2}d and~\ref{panel2}e,
respectively) with the peaks marginally better separated in the
$(u-r)_c$ distribution. We separate the galaxy population into two
classes with cuts at $(u-r)_g=2.1$ and $(u-r)_c=2.35$ respectively.

\subsection{Stellar mass-to-light ratios}
\cite{bell01} provide a standard prescription to determine stellar
masses and stellar mass-to-light ratios from broad band colours. From
their Table 1 we have:

~

$\log(M^*/L_B) = -1.224 + 1.251 (B-R)$.

~

\noindent
From \cite{mgc3} we have the following conversions
from SDSS-DR1 photometry of:

~

$(B-R) = 1.54(g-r)+0.35$, 

~

\noindent
and can recast the above as:

~

$(M^*/L_B) = 10^{[{1.93(g-r)-0.79}]}$,

~

\noindent
where $M^*/L_B$ is the $\bmgc$ stellar mass-to-light ratio in solar
units.  We elected to use the SDSS-DR1 colours for consistency rather
than $(\bmgc-r)$ because the differing deblending algorithms
occasionally lead to spurious ($\bmgc-r)$ colours. Not surprisingly,
as the stellar mass-to-light ratios are derived directly from the
global-$(g-r)$ colour, we again see (Fig.~\ref{panel2}f) a bimodality
with an obvious division at $\log(M^*/L_B)=0.45$.

\subsection{The MGC S\'ersic indices}
The MGC S\'ersic profile indices ($n$), are derived using GIM2D
(\citealp{gim2d}) and the process is described in detail in
\cite{allen05}. Briefly, GIM2D performs a $\chi^2$-minimisation
between the image and a 6-parameter model image. A segmentation mask
derived from SExtractor is used to indicate which pixels are
associated with the object, and an appropriate PSF image is convolved
with the model image. The software uses the standard S\'ersic profile
(\citealp{sersic}, \citealp{graham05}) returning the total flux,
S\'ersic index ($n$), half-light radius ($R_e$), ellipticity, position
angle and positional and background offsets. Fig.~\ref{panel2}g shows
the distribution of the best-fitting S\'ersic parameters for the
MGC. The distribution in $\log(n)$ is bimodal, presumably reflecting
bulge dominated (i.e., pressure supported) or disc dominated (i.e.,
rotationally supported) systems. Note that it is more appropriate to
plot $\log(n)$ rather than $n$, as $n$ appears in the exponent of the
intensity equation, {\it i.e.,} a small change in $n$ makes a large
difference for low $n$ values but minimal difference for high-$n$
values. We adopt a division at the saddle point of $n=2.0$. From
repeat observations of $702$ objects we have
quantified the random error on the S\'ersic index to be $\Delta
\log(n) = \pm 0.041$.

\subsection{The MGC central surface brightnesses}
The (extrapolated) central surface brightness ($\mu_0$) is calculated
directly from the GIM2D S\'ersic only fits and hence strongly
dependent (although not exclusively) on the S\'ersic index.  It is
included here for completeness as many earlier studies have subdivided
the population by a surface brightness cut. The distribution is
bimodal and we divide the galaxy population at $\mu_0 = 19 ~\bmgc$
\mpas. Note that in our previous studies ( \citealp{cross00};
\citealp{cross02}; \citealp{driver05}) we explored the effective
surface brightness absolute magnitude plane. The distinction is
important, whereas bulges and discs typically have similar effective
surface brightnesses their central surface brightnesses are
significantly different. For overcoming selection bias the former is
more relevant, for exploring variations the latter is more appropriate.

\subsection{The MGC half-light radii}
The half-light radius ($R_e$), is calculated directly by the GIM2D
code and is therefore PSF-corrected. The distribution is relatively
broad and peaked at around $10$ kpc. Hence, while bimodalities exist
in many observable properties, the size distribution is smooth (and
hence the effective surface brightness as well). There is no obvious
location at which to divide the population.

\subsection{The observed (uncorrected) distributions}
Fig.~\ref{panel2} shows the {\it observed} distributions, i.e.,
uncorrected for volume bias, for galaxies with $\bmgc < 20$ mag and
$M_{\bmgc} - 5 \log h < -16$ mag, except for morphological type and
$\eta$ which are only known to $\bmgc = 19$ mag.  Note that the
absolute magnitude limit essentially confines this study to the giant
galaxy regime (with some contribution from the brightest dwarf
ellipticals).  In Fig.~\ref{panel1} the overall distributions also shown
subdivided according to MGC continuum type (solid, El; hashed, Sa; or
diagonal lines, Sc). This highlights the correlated nature of the
bimodalities.  However, it is also clear that the subdivisions of the
galaxy population based on colour, morphological type, etc., will each
extract distinct subsets of the galaxy population. Hence, great care
must be taken when comparing high and low redshift samples subdivided
by different methods. Fig.~\ref{panel1} shows the same distributions
but restricted to the $\bmgc < 19$ mag population. The morphological
groups (E/S0, Sabc, Sd/Irr) are shown with distinctive shading
throughout (solid, hashed, diagonal lines, respectively). From
Fig.~\ref{panel1} it is clear that the E/S0 (bulge-dominated) and
Sd/Irr (disc-only) populations occupy the extremes of the bimodal
distributions with the Sabc distribution as a bridging population.

\begin{figure*}
\centering\includegraphics[width=\textwidth]{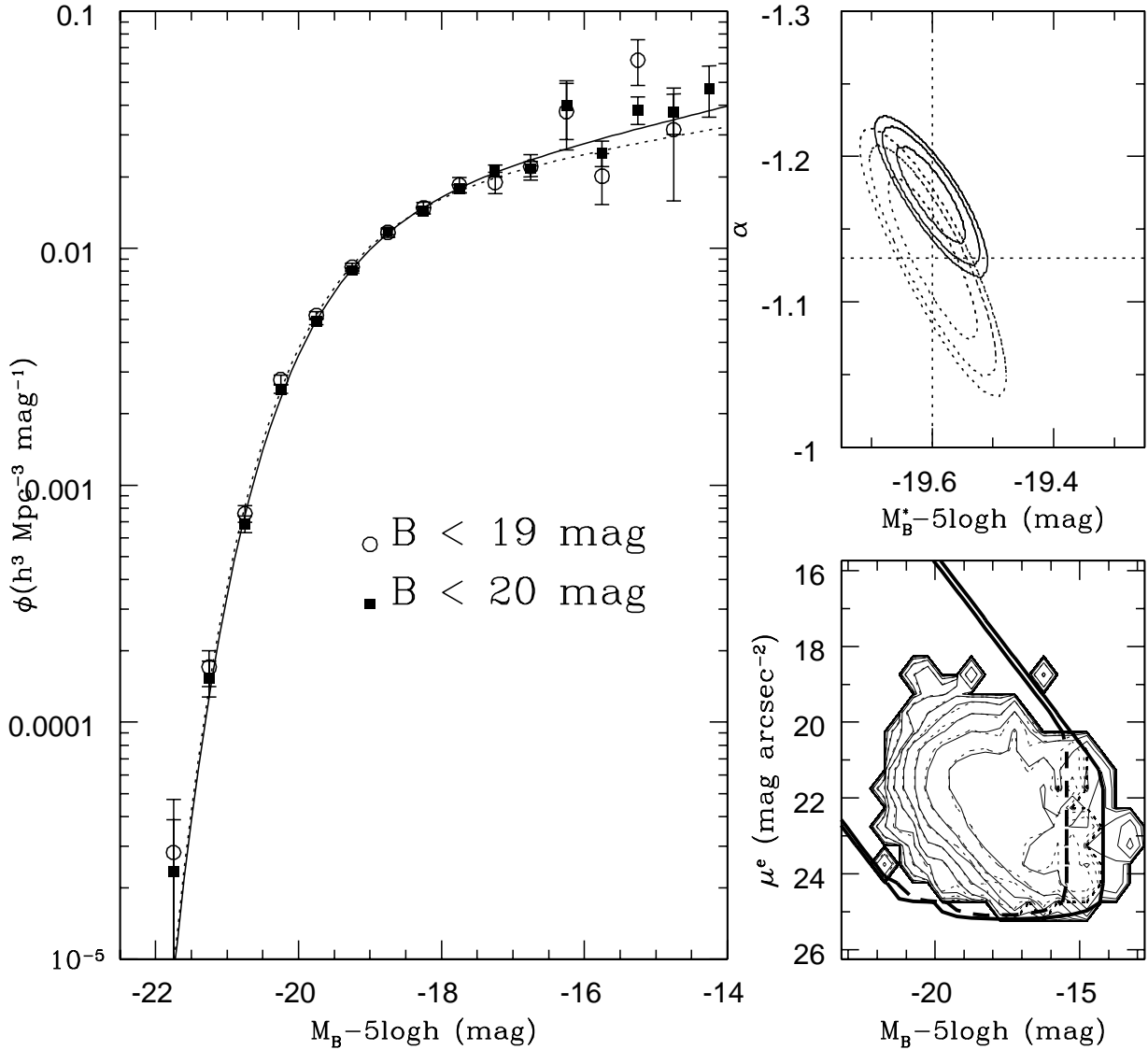}
\caption{({\it main panel}) the global $B$-band luminosity function
derived via our modified step-wise maximum likelihood method for the
$\bmgc < 19$ (solid squares and solid line) and $\bmgc < 20$ (open
circles and dotted line) mag samples. ({\it upper right}) the 1,2 and
3 $\sigma$ error-ellipses of the Schechter function fits and ({\it
lower right}) the luminosity surface brightness plane and the
logarithmic contours for the $\bmgc < 19$ (dotted line) and $\bmgc <
20$ mag (solid line) samples. The two dashed lines denote the limits
of reliability for the two samples.}
\label{newlfs}
\end{figure*}

\section{The luminosity distributions and Schechter functions}
To obtain luminosity distributions for sub-populations we first
re-derive the global bivariate brightness distributions for the
$\bmgc < 19$ mag and $\bmgc < 20$ mag samples following
\cite{driver05}. The method implemented in \cite{driver05} is a
variant of the standard step-wise maximum likelihood (SWML; see
\citealp{eep}) operating in the plane of luminosity {\it and} surface
brightness (as opposed to luminosity alone). The main advantage is the
ability to track the five key selection boundaries (see
\citealp{driver99}) which govern whether a galaxy is reliably detected
or not: maximum/minimum luminosity, maximum/minimum size, and maximum
mean effective surface brightness. The analysis\footnote{In detail a
few minor changes have occurred in the analysis since
\cite{driver05}. Firstly the selection boundaries are implemented {\it
before} the seeing correction; secondly the incompleteness matrix is
also derived prior to seeing correction; and thirdly the normalisation
is now made to galaxies in the range: $-20.0 < M_{\bmgc} < -19.5$
mag and $20.75 < \langle \mu_{\mbox{eff}} \rangle < 23.25$ \mpas and
$0.047 < z < 0.162$. Minor improvements to the code have also resulted
in a more accurate evaluation of LSP cells which are intersected by
the selection limits. These updates do not change the results of
\cite{driver05} significantly.}  assumes 
($\Omega_{M} = 0.3, \Omega_{\Lambda} = 0.7$) with $H_{o}=100$h km
s$^{-1}$Mpc$^{-1}$ and K-corrections derived individually for each
galaxy (see \citealp{driver05} for specific
details). Fig.~\ref{newlfs} shows a comparison of the luminosity
distributions, functions, error ellipses and LSP distributions for
these two samples which are fully consistent with each other.

To derive the subdivided luminosity distributions, $\psi(M,x)$ we now
simply calculate: 
\begin{equation}
\psi(M,x)=\sum_{\mu=-\infty}^{\mu=\infty}[f_x(M,\mu).\phi(M,\mu)] \delta\mu,
\end{equation}
where $f_x(M,\mu)$ represents the apparent frequency of some
attribute, $x$, (e.g., fraction of E/S0 type) within that absolute
magnitude and surface brightness bin, $\delta M \delta \mu$.
$\phi(M,\mu)$ represents the space density of all galaxies in that
same interval.  The integral over surface brightness then converts the
bivariate distribution to the monovariate luminosity function; note
that although the integral ranges across all values of surface
brightness the distribution at any $M$ is actually bounded by the
selection limits. This is essentially identical to re-deriving the LSP
(also known as the bivariate brightness distribution or BBD, see
\citealp{phillipps95}) for each sub-population but computationally
simpler and more stable.  Figs.~\ref{morpha}--\ref{sersic} show the
recovered luminosity distributions for the various subdivisions.  In
each case the luminosity distributions were fitted by a Schechter
function to $M_{\bmgc} -5 \log h = -16$ mag. In each figure the
recovered data are shown (left) along with the best Schechter function
fits and the 1-,2- and 3- $\sigma$ error contours (right-side). The
fitting parameters are also tabulated in Table~\ref{table2} along with
the 1-$\sigma$ errors from the $\chi^2$ fit, the integrated
luminosity density ($j_{b_j}$), the median $(u-r)_g$ and the total
stellar mass for each subdivision. The luminosity densities and stellar
densities $(M^*)$ were derived as follows:

~

$j_{b_j}=\phi^*10^{-0.4(M_B-M_{\odot})}\Gamma (\alpha+1)$, 

~

\noindent and,

~

$M^* = \sum_i^N 10^{[{1.93(g-r)-0.79})}10^{-0.4(M_B-M_{\odot})}$.

~

\noindent
i.e., analytically for the luminosity density and an empirical sum for
the stellar mass density because of the strong colour dependency.

\subsection{Luminosity distributions/functions by morphological type}
\begin{figure*}
\centering\includegraphics[width=\textwidth]{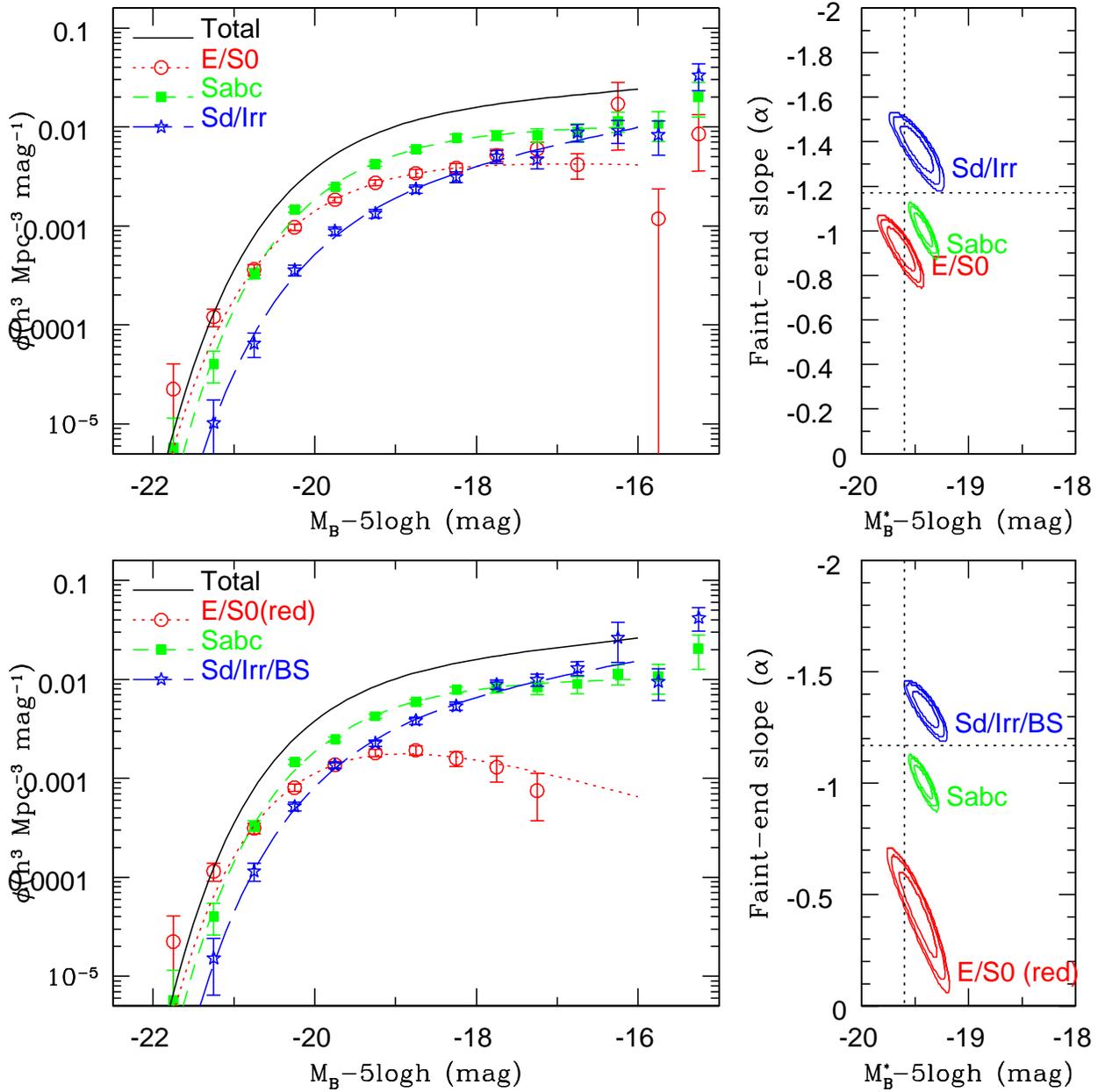}
\caption{({\it upper}) The derived data and fitted Schechter
functions subdivided by morphological type and ({\it lower}) with the blue 
spheroids (BS) moved into the Sd/Irr category.}
\label{morpha}
\end{figure*}

In Fig.~\ref{morpha} (upper) we show the eyeball morphological
luminosity distributions/functions. Although the three divisions show
segregation in their Schechter function parameters, the distinction
between the E/S0 and Sabc categories is not particularly
strong. Certainly the E/S0s do not appear to follow the Gaussian
distribution anticipated by \cite{bst88}, reported in \cite{jerjen},
and discussed extensively in \cite{lapparent}. In Ellis et al. (2005,
their fig.~3) we found that our eye was often fooled by smooth, blue,
low-luminosity spheroidal systems. This highlighted both the
subjective nature of eyeball classification and its susceptibility to
variations in data quality. Whether the blue spheroid population
constitutes the tail-end of the 'downsizing' phenomenon or interlopers
is at this stage unclear but under investigation. If one uses
additional colour information to separate out the blue and red
spheroids (i.e., following Ellis et al.)  and re-derives the
morphological luminosity distributions we now get the distributions
shown in Fig.~\ref{morpha} (lower). The luminosity distribution of the
blue spheroid (BS) population alone (not shown but see
Table~\ref{table2}) follows almost exactly the same distribution shape
as the Sd/Irr population and most likely has a common origin.  This is
also consistent with the idea that the E/S0 population has indeed been
contaminated by a population of smooth blue dwarf systems (dwarf
spheroidals). In so far as a comparison can be made, our values could
be contrasted to those listed in Table 4 of \cite{lapparent} where we
see measurements for the E/S0 class from the Stromlo APM survey
($M^*_{\bmgc}=-19.33$ mag, $\alpha=+0.2$; \citealp{loveday}), the
Southern Sky Redshift Survey 2 ($M^*_{\bmgc}=-19.64$ mag,
$\alpha=-1.00$; \citealp{ssrs2}) and the Nearby Optical Galaxy Sample
($M^*_{\bmgc}=-19.80$ mag, $\alpha=-0.97$; \citealp{nog}). These three
surveys concur well with our uncontaminated and contaminated E/S0
samples suggesting the problem of BS contamination may be endemic in
all eyeball classified samples (and hence in ANN classification as
well) -- see \cite{ellis} for the formal detection of these systems
and also \cite{cross04} for discussion of BS contamination of field
ellipticals at $z \sim 0.75$ (r.f., fig.2a from \citealp{bell05}).
While arguably a combined eye and colour cut may provide a more robust
division, the process of morphological classification (whether by eye
or ANN) is subjective and hence susceptible to confusion,
controversy and ambiguity.

\subsection{Luminosity distributions/functions by $\eta$ and MGC continuum type}
\begin{figure*}
\centering\includegraphics[width=\textwidth]{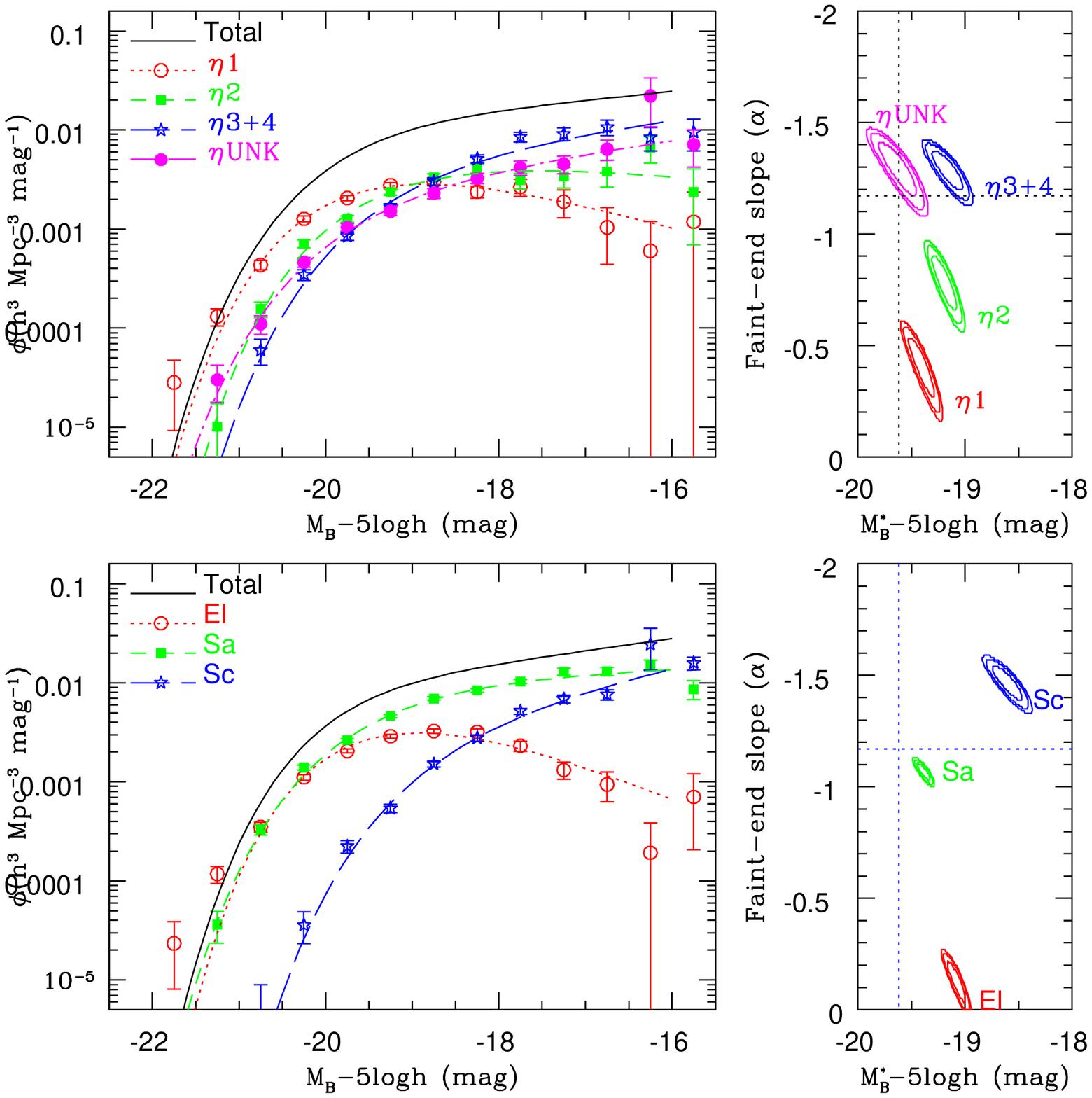}
\caption{({\it upper}) The derived data and fitted Schechter
functions subdivided by 2dFGRS-$\eta$ parameter and ({\it lower}) subdivided
by MGC continuum type.}
\label{seta}
\end{figure*}

Fig.~\ref{seta} shows the luminosity distributions/functions
subdivided by $\eta$ (upper) or MGC continuum type (lower).  The
luminosity distributions/functions subdivided by $\eta$ type concur
with the earlier results by the 2dFGRS team (i.e., \citealp{folkes};
\citealp{madgwick}), with the $\eta 1$-type showing a distinct luminosity
distribution and less distinction between the remaining $\eta$
types. Unfortunately the fraction of galaxies without a known $\eta$
type is significant (even to $B < 19$ mag) due to the requirement for
moderately good signal-to-noise spectra. 

Our continuum typing, based on spectral template fitting to the
available $uBgriz$ broad-band photometry, appears significantly more
promising with the three continuum shapes (see
Fig.~\ref{spectraltypes}) represented by very distinct luminosity
distributions/functions. From Fig.~\ref{panel2} we see that these
distinctions (shown as shaded, hashed and lines histograms) are well
correlated with other observables. While correlation between continuum
type and colour (and hence stellar mass-to-light) is to be expected,
the correlation with structural measurements is significant as no
structural information was used in the determination of the MGC
continuum type. The El continuum type extracts the red, high
mass-to-light ratio, concentrated, and predominantly luminous
systems. The Sa class exhibits a broader range bridging the El and Sc
populations. The Sc class constitutes the bluest and least
concentrated systems. The implication is that the El systems are
luminous (massive) and old (or at least their stars are), and the Sc
systems appear to be predominantly low luminosity (low mass), and
young (late-formers).

\subsection{Luminosity distributions/functions by colour and mass-to-light ratio}
\begin{figure*}
\centering\includegraphics[width=\textwidth]{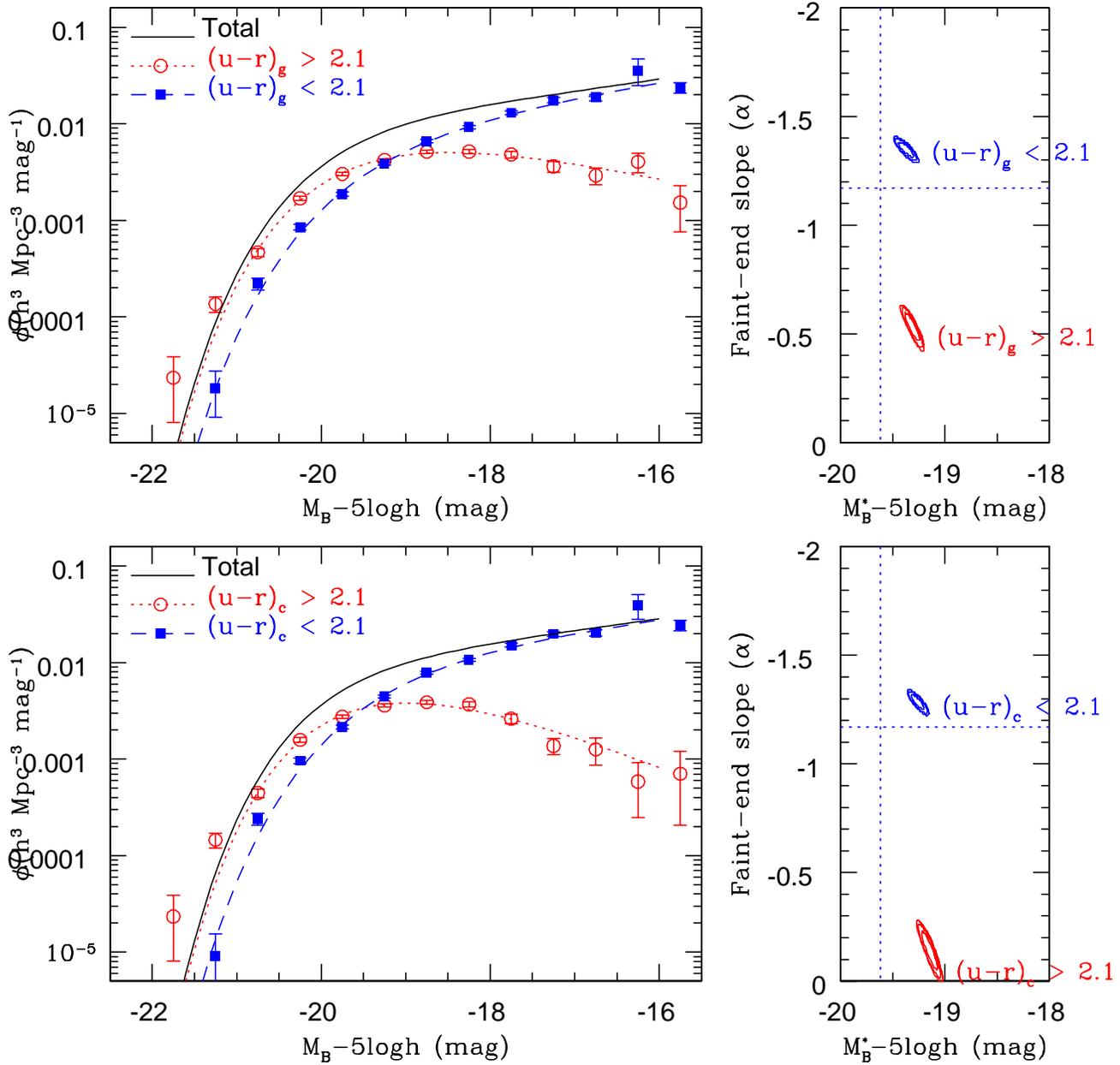}
\caption{The derived data and fitted Schechter
functions subdivided by $(u-r)_g$ colour ({\it upper}) and $(u-r)_c$
colour ({\it lower})}
\label{colour}
\end{figure*}

Fig.~\ref{colour} shows the luminosity distributions/functions
subdivided by $(u-r)_g$ (upper) and $(u-r)_c$ colours (lower).  It is
worth noting that our calculation of the rest colour is independent of
the standard SDSS method, based on our own spectral template fitting
routine which includes all 27 spectral templates given by
\cite{poggianti97}, see \cite{driver05} for details. The luminosity
distributions of the red and blue populations are significantly
distinct. Worth noting is that the $(u-r)_c$ colour -- derived from
SDSS-DR1 PSF magnitudes -- appears to segregate galaxies better than
the global colour -- derived from SDSS-DR1 Petrosian magnitudes. The
improvement in segregation (see Fig.~\ref{colour} right side panels)
is significant and could be interpreted as better separating the
galaxy population into those with dominating old bulges and those
without. The key question is why do red bulges predominantly occur in
luminous galaxies?  Presumably either the low-luminosity bulges are
unresolved and hence hidden within the PSF, or that old inert (and
therefore red) low-luminosity bulges are rare\footnote{Although dwarf
ellipticals are considered the natural extension of the disc-less
spheroids, \citealp{graham03}, there is no obvious low-luminosity
bulge-disc counterpart.}. This is consistent with some studies which
argue that the bulges of late-type systems are pseudo-bulges built up
through inner disc and bar instabilities (e.g., \citealp{erwin})
rather than genuine bulges.  As genuine bulges should be cuspier one
might expect the core colour to be better at identifying such
systems. As mentioned in Section 2.4 our core colour is not ideal but
simply relies on the colour within the smallest available aperture. An
obvious extension would be to derive the core colour from multicolour
2D bulge-disc decomposition models to see whether the red peak becomes
even sharper.

The SDSS $(u-r)_g$ colour bimodality was studied in detail by
\cite{baldry} who derived the $r$-band luminosity functions and
identified faint-end parameters of $\alpha = -1.35\pm0.05$ and
$\alpha=-0.83\pm0.002$ for the blue and red populations
respectively. This compares with our values of $\alpha=-1.35\pm0.02$
and $\alpha=-0.54\pm0.05$ for our $B$-band selected samples. As one
expects a $B$-band sample to preferentially miss red systems when
compared to an $r$-band sample, the higher $\alpha$ for the red
systems is unlikely to be significant. As the mass-to-light ratios are
derived from the global $(g-r)$ colour (via \citealp{bell01}) the
narrowness of the mass-to-light ratio distribution is
expected. However, the MGC continuum classification appears remarkably
robust at identifying systems occupying the high mass-to-light ratio
peak (Fig.~\ref{panel2}f). Perhaps even more interesting is the extreme
narrowness of the high-M/L peak, essentially a representation of the
well known colour-magnitude relation of cluster ellipticals
(\citealp{sv78}). This narrowness presumably reflects the universal
endpoint of galaxy evolution, which in turn places a joint constraint
on the age, metallicity and even the universality of the IMF (see
\citealp{bc03}).

\subsection{Luminosity distributions/functions by structure}
\begin{figure*}
\centering\includegraphics[width=\textwidth]{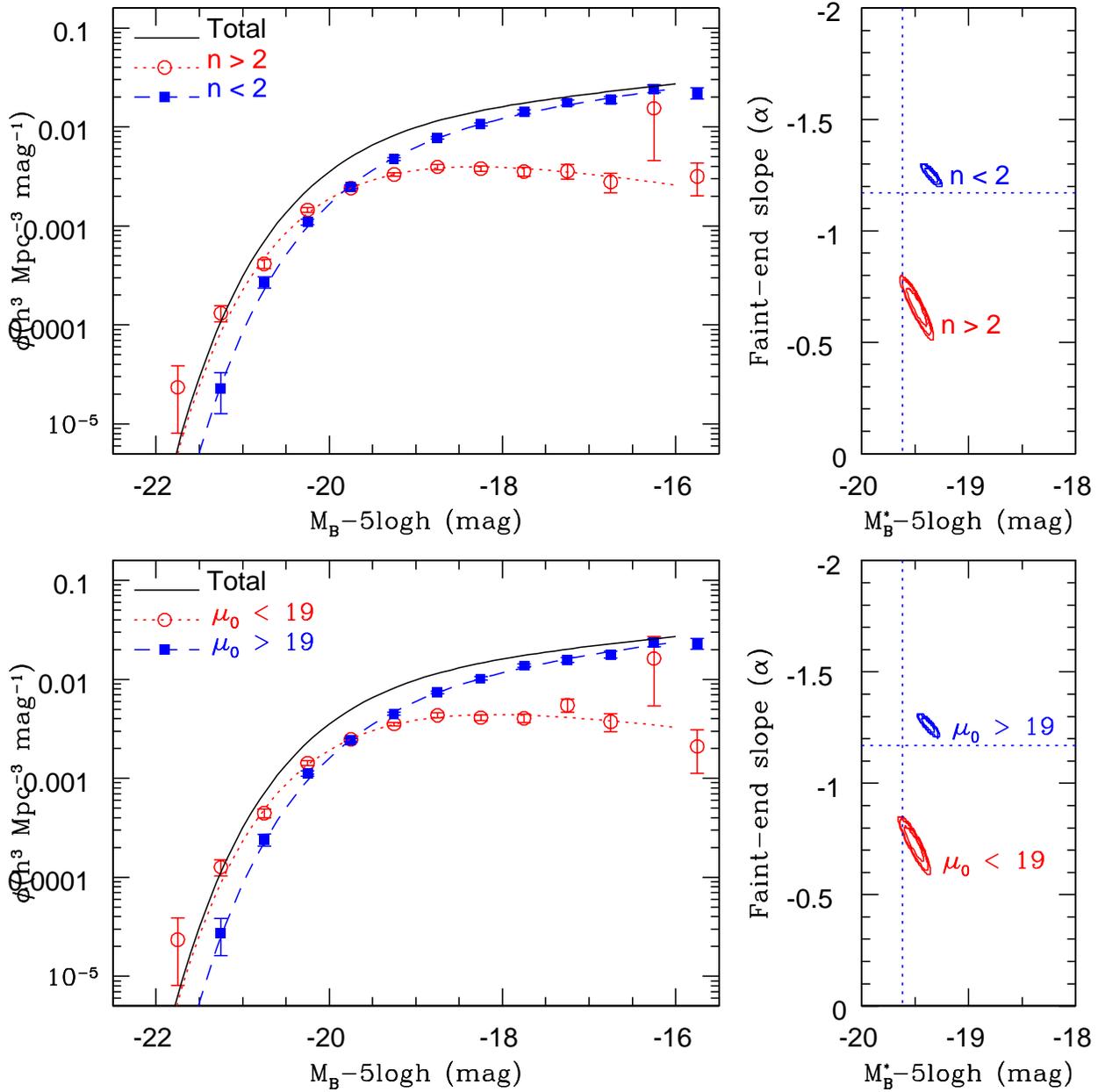}
\caption{The derived data and fitted Schechter
functions subdivided by $\log ($S\'ersic index$)$ ({\it upper}) and fitted
central surface brightness ({\it lower}).}
\label{sersic}
\end{figure*}

Fig.~\ref{sersic} shows the luminosity distributions/functions
subdivided by S\'ersic index, $n$ (upper), and central surface
brightness (lower).  As the central surface brightness is derivable
from $n$, total magnitude and effective radius, it is not independent
and hence the similarity in the two distributions (upper and lower
panels) is to be expected. Taking the $\log(n)$ division we see that
while the high and low-$n$ populations follow distinct luminosity
distributions, the distinction is not as great as for the MGC continuum
and colour divided distributions.  This is, at least in part, due to
the fact that some dEs with $-16 > M_{\bmgc} - 5 \log h > -18$ mag
will have $n < 2$ and others will have $n > 2$. Folding in colour
information to separate the red dEs from the blue discs will help and
is considered in the following section.

\subsection{Luminosity distribution summary}
The main conclusions from this section are: 

\noindent
(1) Morphological classification based on the eye, including
Artificial Neural Networks which use a training set, are fraught with
difficulty because of the subjective nature of the classification
process. 

\noindent
(2) In so far as morphological luminosity functions can be obtained,
particular care must be taken to distinguish genuine ellipticals from
smooth, blue, low-luminosity systems and eyeball classifications
should be made using colour images if possible.

\noindent
(3) Both global and core $(u-r)_{g,c}$ colours segregate the population
well into blue and predominantly luminous red systems.

\noindent
(4) The core colour appears to better segregate the two populations
 (in $M-\alpha$-space) than the global colour, probably due to the
 blending of colours from red bulges embedded in blue discs.

\noindent
(5) The narrowness of the red, core-colour peak {\it may} imply an
early formation age from a universal initial mass function for these
systems (as commonly assumed, e.g., \citealp{bell01}).

\noindent
(6) Further gains {\it may} be made by moving beyond a single colour
cut to a continuum-type classification system based on multiple
colours. This highlights three distinct populations. A luminous red
population a low luminosity blue population and an intermediate
population, these are closely associated with E/S0 (bulge-dominated),
Sd/Irr (disc-only) and Sabc (bulge and disc) systems.

\noindent
(7) Galaxies can also be effectively separated into two populations
    using structural division in $\log(n)$ space. The populations
    constitute a concentrated and diffuse population.

\noindent
(8) Colour appears to segregate the galaxy population more effectively
than any other single measurement (i.e., S\'ersic index or
surface brightness).

~
\noindent
In the next section we continue the analysis by exploring selected
bivariate distributions.

\section{Bivariate distributions}
We derive volume corrected bivariate distributions/functions in a
similar manner to the luminosity distributions by starting from the
overall bivariate brightness distributions for the entire $\bmgc < 19$
and $\bmgc < 20$ mag population and {\it projecting} attributes $x$
and $y$ on to a new plane, i.e.,
\begin{equation}
\psi(x,y)=\sum_{M_{\bmgc}=-\infty}^{M_{\bmgc} < -16} ~~ \sum^{\mu=+\infty}_{\mu=-\infty}f_{x,y}(M,\mu).\phi(M,\mu) \delta M \delta \mu,
\end{equation}
where the values are defined as for
Eqn.~(1). Fig.~\ref{colourx}--\ref{hlrx} show the bivariate
distributions of: $(u-r)_g$, $(u-r)_c$, $\log (n)$, $\mu_0$
and $\log(R_e)$ versus $M_{\bmgc}$. For each figure we show (upper
left) the raw observed distribution as both data points and contours,
and ({\it lower left}) the volume corrected distributions using
Eqn.~(2). The right-side panels show the histograms along the absolute
magnitude axis. As a reminder, only galaxies with $M_{\bmgc} -5 \log h
< -16$ mag are included in the analysis.

\subsection{Bivariate distributions by colour}
\begin{figure*}
\centering\includegraphics[width=\textwidth]{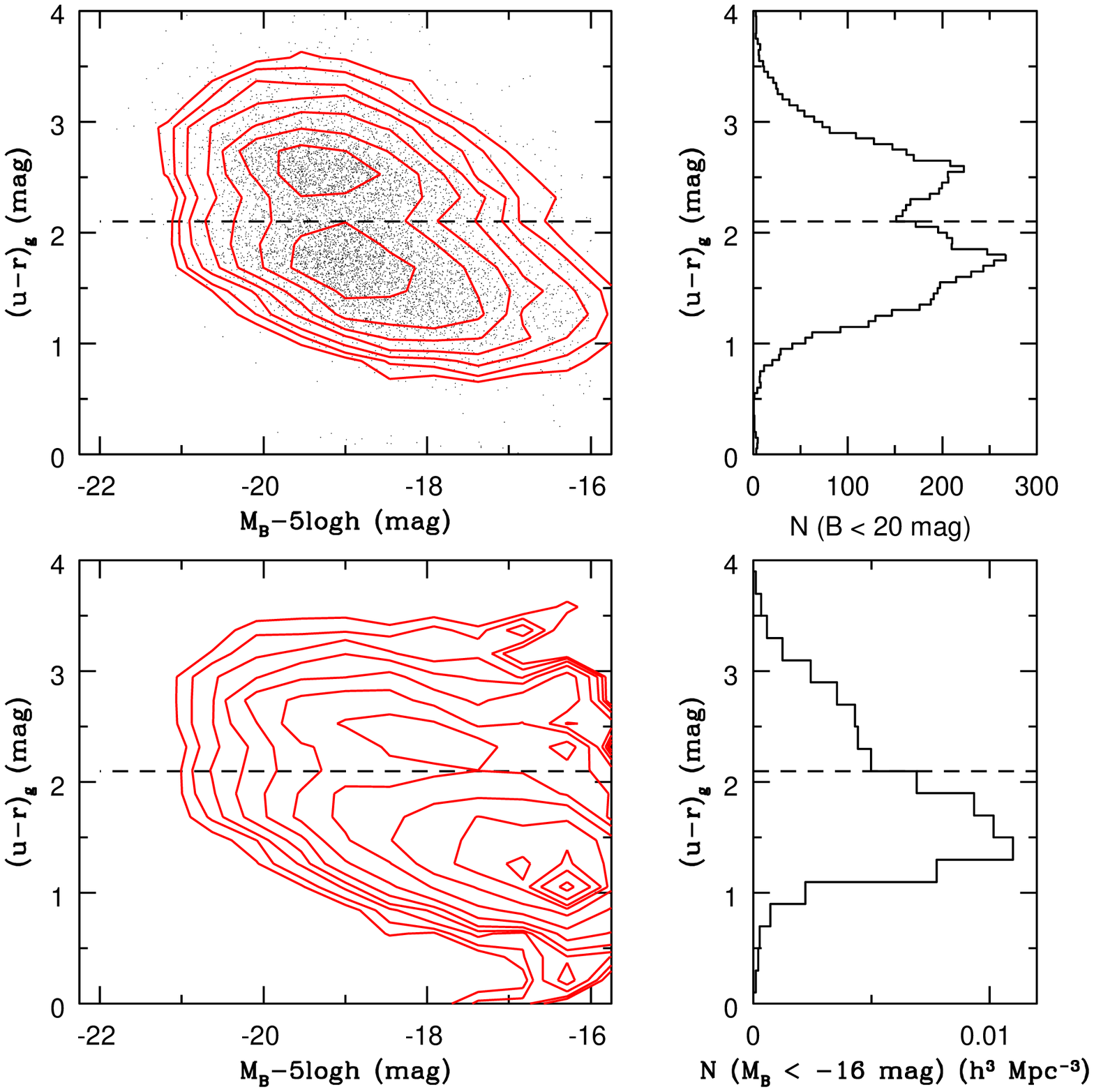}
\caption{({\it upper left}) the observed $(u-r)_g$ versus $M_{\bmgc}$
distribution (dots and 0.2 dex contours) and histogram ({\it upper
right}). ({\it lower left}) the volume-corrected $(u-r)_g$ versus
$M_{\bmgc}$ distribution (dots and 0.2 dex contours) and histogram ({\it
lower right}).}
\label{colourx}
\end{figure*}

\begin{figure*}
\centering\includegraphics[width=\textwidth]{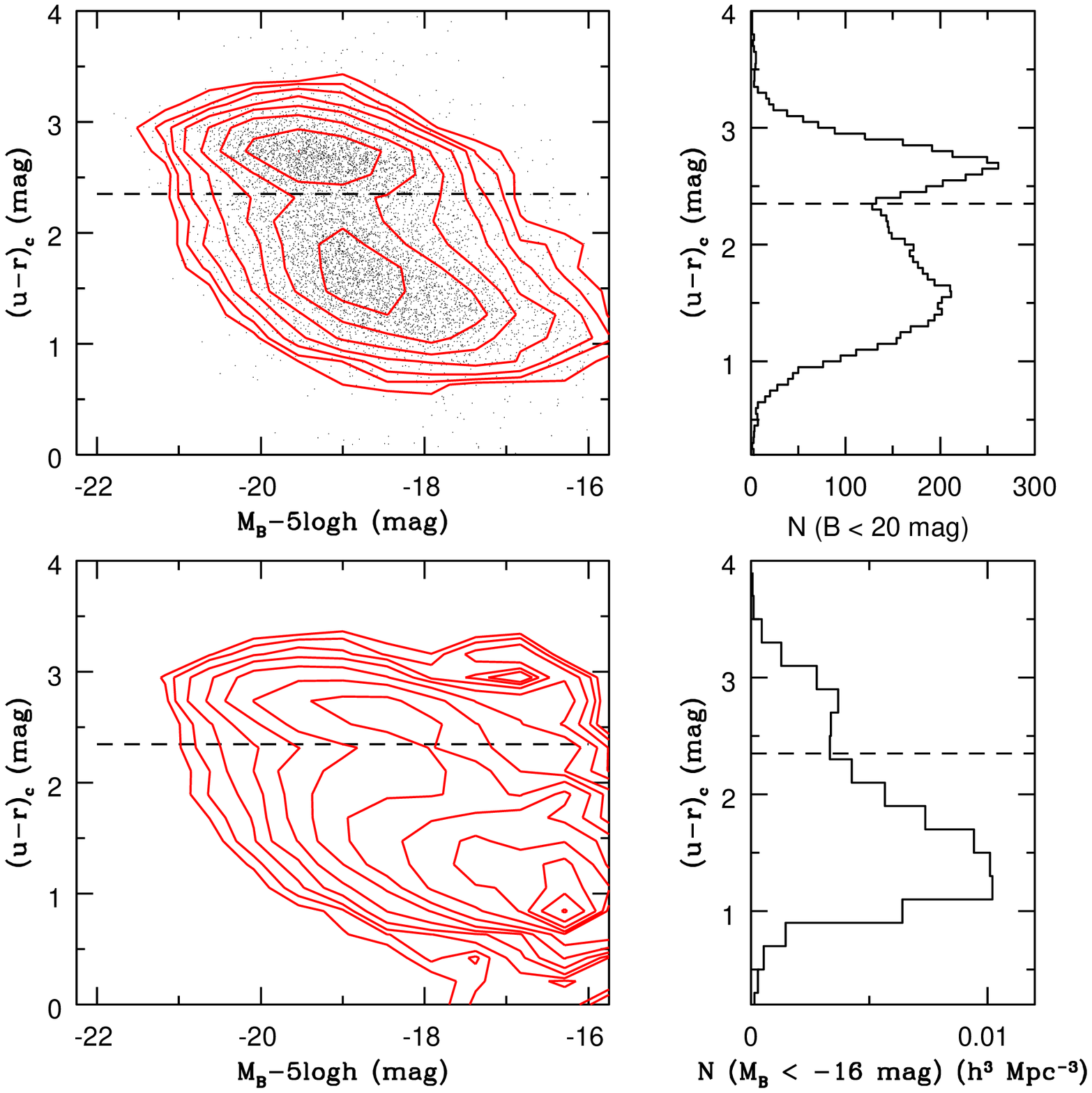}
\caption{({\it upper left}) the observed $(u-r)_c$ versus
$M_{\bmgc}$ distribution (dots and 0.2 dex contours) and histogram
({\it upper right}). ({\it lower left}) the volume-corrected $(u-r)_c$
versus $M_{\bmgc}$ distribution (dots and 0.2 dex contours) and
histogram({\it lower right}).}
\label{corecolourx}
\end{figure*}

Figs.~\ref{colourx} and ~\ref{corecolourx} show the bivariate
distributions of rest $(u-r)_g$ and $(u-r)_c$ colours versus
$M_{\bmgc}$. In both figures we see a distinct red and blue sequence
as characterised by \cite{baldry} using SDSS data. In both figures the
red distribution appears entirely bounded albeit overlapping with the
luminous end of the more extensive blue population. Both the red and
blue sequences show colour gradients reminiscent of the
luminosity-metallicity relation seen in nearby rich clusters. Note
that bimodality is only obvious in the observed histograms and not in
the volume corrected histograms as the numbers are dominated by the
lower luminosity systems. In terms of stellar mass (not shown) the
bimodality remains but is skewed towards the red peak. The division
between the red and blue populations is more distinct (and the red
peak histogram narrower) when $(u-r)_c$ colour is used. This implies
that the core colour is the more fundamental (less contaminated)
measure. Further data probing to lower luminosities is required to
unequivocally conclude that the red sequence is bounded at low
luminosity (i.e., few galaxies with $(u-r)_c > 2$ and $M_{\bmgc} -5
\log h < -17$ mag). An indication of a low luminosity red spike (dEs ?)
is seen but with extremely low significance because of limited
statistics and the large volume amplification at the faint-end.

\subsection{Bivariate distributions by structure}
\begin{figure*}
\centering\includegraphics[width=\textwidth]{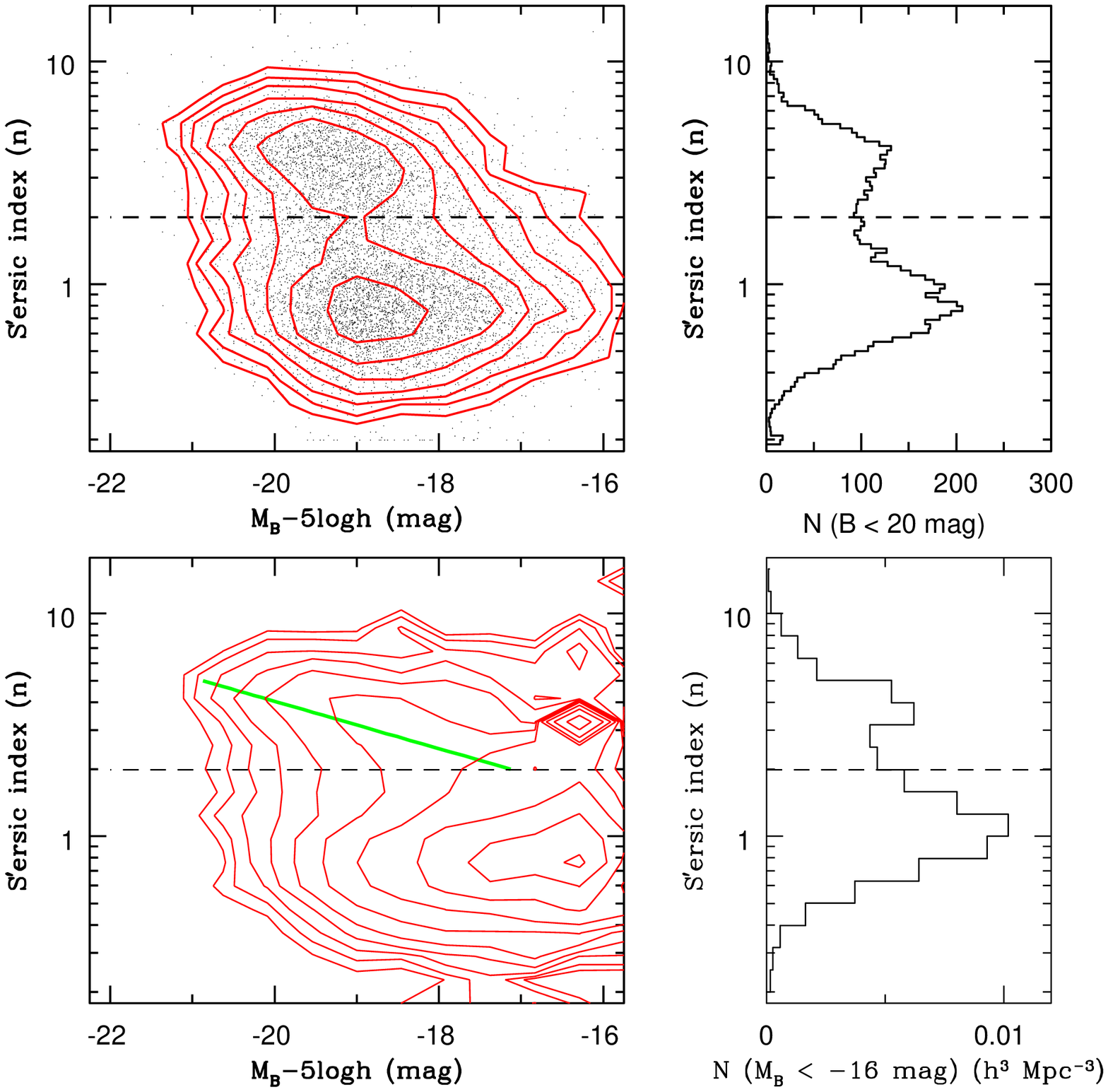}
\caption{({\it upper left}) the observed $\log(n)$
versus $M_{\bmgc}$ distribution (dots and 0.2 dex contours) and histogram
({\it upper right}). ({\it lower left}) the volume-corrected
$\log(n)$ versus $M_{\bmgc}$ distribution (dots and 0.2 dex
contours) and histogram ({\it lower right}).  The solid line shows the
known relation for spheroids (see Graham \& Guzm\'an 2003).}
\label{sersicx}
\end{figure*}

\begin{figure*}
\centering\includegraphics[width=\textwidth]{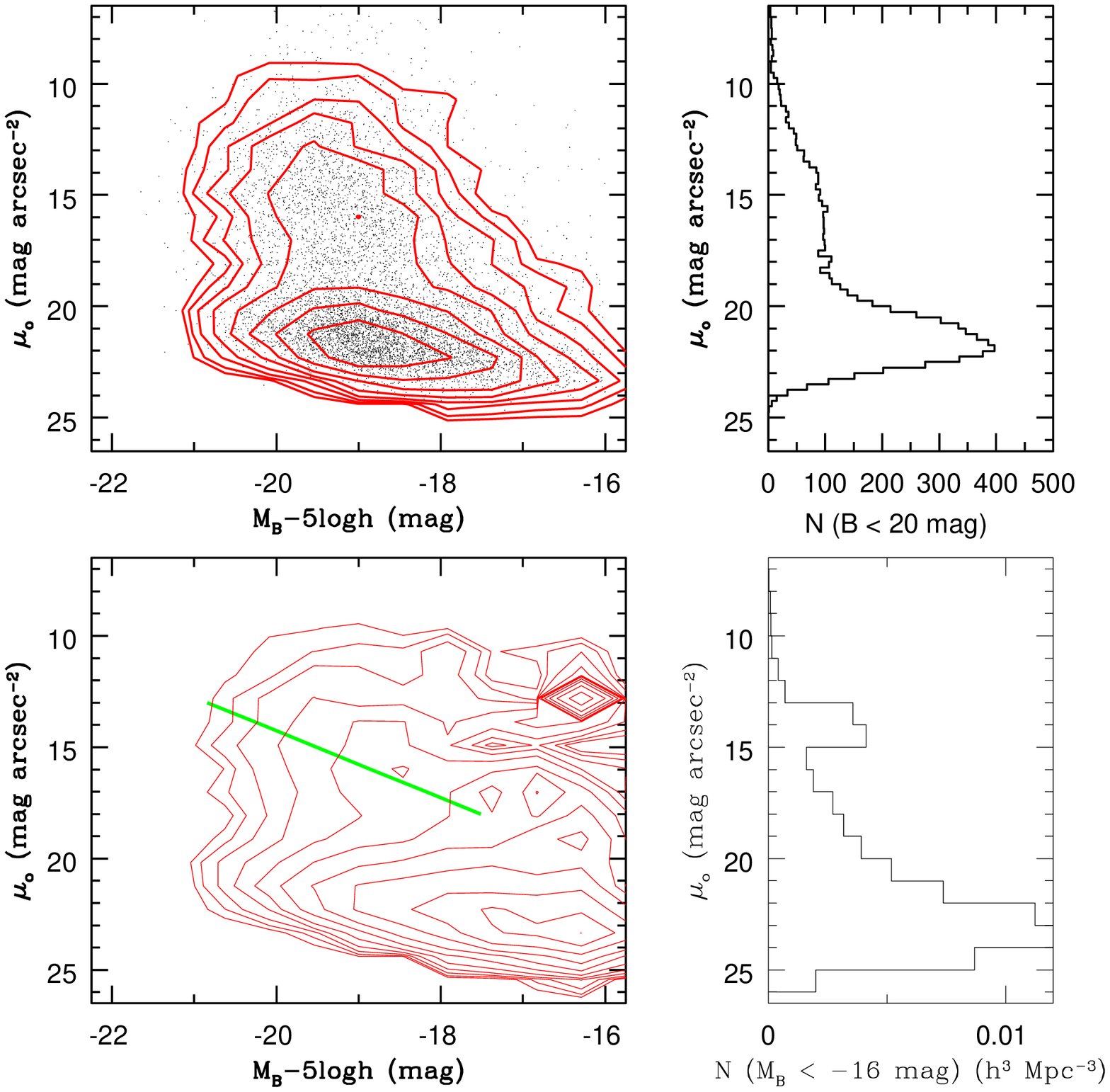}
\caption{({\it upper left}) the observed $\mu_{o}$ versus $M_{\bmgc}$
distribution (dots and 0.2 dex contours) and histogram ({\it upper
right}). ({\it lower left}) the volume-corrected $\mu_{o}$ versus
$M_{\bmgc}$ distribution (dots and 0.2 dex contours) and histogram ({\it
lower right}).  The solid line shows the known relation for spheroids
(see Graham \& Guzm\'an 2003).}
\label{sbx}
\end{figure*}

\begin{figure*}
\centering\includegraphics[width=\textwidth]{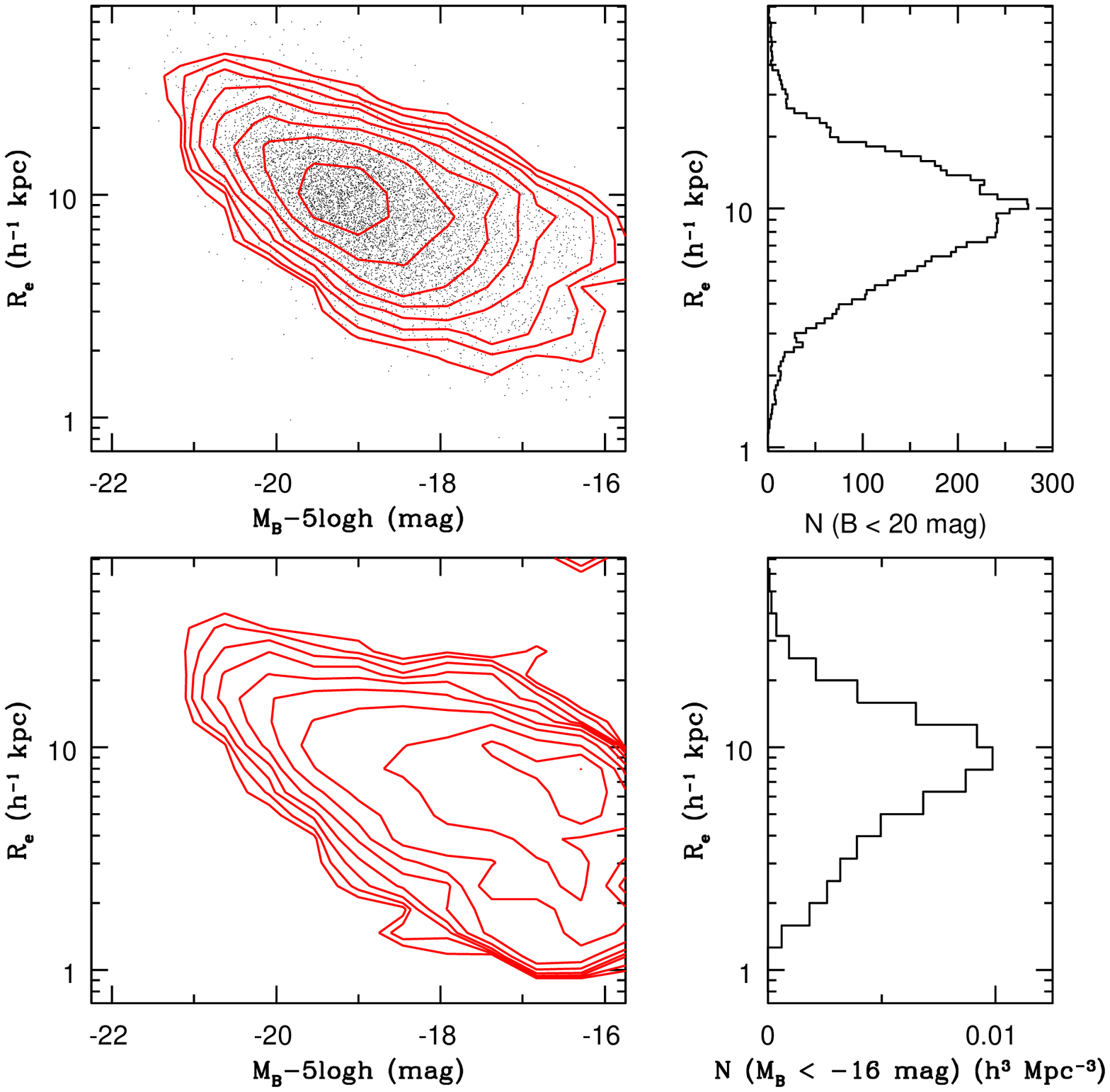}
\caption{({\it upper left}) the observed HLR versus $M_{\bmgc}$ distribution
(dots and 0.2 dex contours) and histogram ({\it upper right}). ({\it lower left}) the volume-corrected
HLR versus $M_{\bmgc}$ distribution (dots and 0.2 dex contours) and histogram ({\it lower right}).}
\label{hlrx}
\end{figure*}

Figs.~\ref{sersicx},~\ref{sbx} and~\ref{hlrx} show the corresponding
structural bivariate distributions for $\log(n)$, $\mu_0$ and $R_e$
respectively. The first two show bimodality albeit less obviously than
for the global and core-colour distributions. Both Figs.~\ref{sersicx}
and \ref{sbx} suggest a concentrated, high surface brightness,
low-luminosity population close to the limit of analysis (in the
volume corrected distributions). Whether this represents a third
distinct population or a manifestation of the limiting statics close
to the selection boundaries is difficult to establish and requires
additional data to clarify.  One clear possibility is that it may
represent the onset of nucleated dwarf systems which would exhibit
both higher than typical S\'ersic indices and correspondingly higher
central surface brightnesses. Unlike the colour bivariate
distributions, the distinct concentrated giant (E) distribution blends
into the diffuse (low-$n$, disc) population towards lower
luminosities. The size distribution shows no bimodality but a
broadening towards lower luminosity as found by \cite{shen03} and
\cite{driver05}.

\subsection{The joint distribution of core-colour and S\'ersic index}
\begin{figure*}
\centering\includegraphics[width=\textwidth]{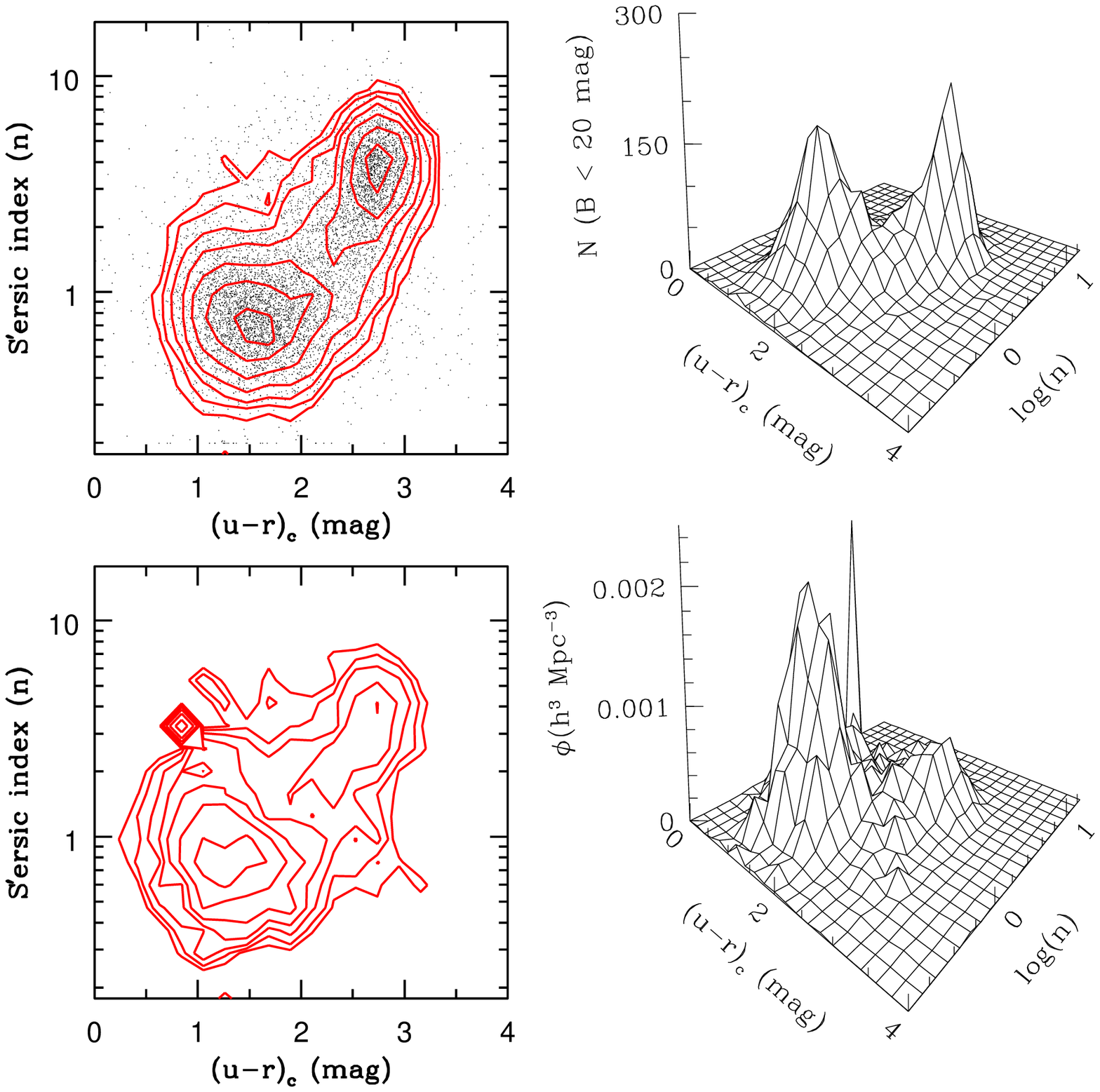}
\caption{(upper) the observed distribution of log(n)
versus $(u-r)_c$ colour (left, shown as data points and 0.2 dex
contours; right shown as a 3D plot with the z-axis linear). (lower)
the volume corrected distribution of $\log(n)$ vs $(u-r)_c$ colour
(left, shown as 0.2 dex contours and right as a 3D plot with z-axis
linear).}
\label{colnx}
\end{figure*}

\begin{figure*}
\centering\includegraphics[width=\textwidth]{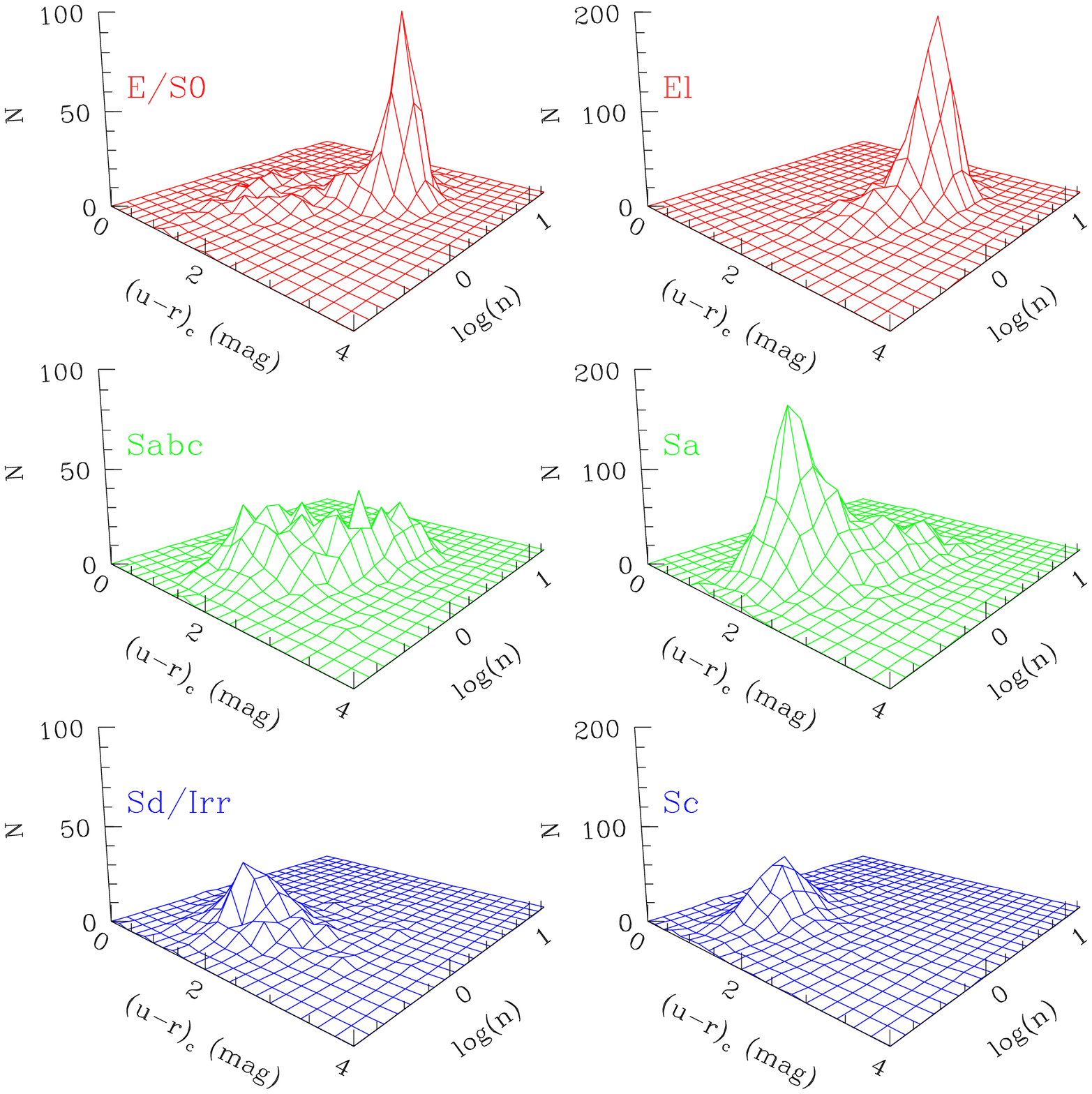}
\caption{(left side) The distribution of E/SO (upper),
Sabc (centre) and Sd/Irrs (lower) in the $\log(n)$ vs $(u-r)_c$
colour. (right side) The distribution of el (upper), Sa (centre) and Sc
(lower) in the $\log(n)$ vs $(u-r)_c$ colour}
\label{colnxx}
\end{figure*}

Having established that bimodality exists in both the colour and
$\log(n)$ distributions, and that the $(u-r)_c$ shows the bimodality
most strongly, we now explore the joint $(u-r)_c$-$\log(n)$
plane. Fig.~\ref{colnx} shows the observed distribution (upper) in
contour form (left) and as a 3D projection (right). The lower panels
show the equivalent for the volume corrected distribution.
Fig.~\ref{colnx} shows two distinct peaks with a bridge between them
pointing towards either two distinct populations or two dominant
characteristics. The peaks are typified by a high S\'ersic index red
peak and a low S\'ersic index blue peak. Note that the additional
third spike (in the lower right panel) may constitute a compact blue population
reminiscent of the blue spheroids (see Section 3.1). The key question
at this stage is whether we are seeing two distinct galaxy populations
(red and blue) or whether the peaks correspond to two characteristics
(i.e., bulges and discs). Bulge-disc decomposition is clearly
essential to establish this unambiguously. However early
indication comes from the morphological classifications where, by
definition (see Section 2.1), E/S0s represent bulge-dominated systems,
Sabcs joint bulge-disc systems and Sd/Irrs disc-only systems.
Fig.~\ref{colnxx} (left) shows the raw observed distributions (i.e.,
equivalent to Fig.~\ref{colnx} [upper right]) for the E/S0 (upper),
Sabc (middle) and Sd/Irr (lower) populations. We see quite clearly
that the bulge-dominated (E/S0) and disc only (Sc) systems lie almost
exclusively in one peak or the other.
The bulge+disc systems (Sabcs) straddle both peaks and the
divide. This is extremely strong evidence that points towards the
fundamental nature of bulges and discs and the need to interpret
colour bimodality as a manifestation of two components rather than two
distinct galaxy populations.  Note that we have shown the original
E/S0 classifications without the additional colour tinkering described
in Section 3.1 and the BS population shows up as fairly distinct and
lying within the blue peak.
Fig.~\ref{colnxx} (right) shows the observed distributions for the
MGC continuum classifications of El (upper), Sa (middle) and Sc (lower).
This suggests that while the El--Sa division is
fundamental the Sa--Sc division is probably not, but instead merely
an arbitrary cut of a continuous distribution.

\section{Discussion}
Our main result is that bimodality occurs in both the rest-$(u-r)$ and
$\log(n)$ distributions and Fig.~\ref{colnx} highlights the 2D nature
of this bimodality with two peaks of approximately equal height and
shape in the observed ($M_{\bmgc} < -16$ mag) plane (upper). We have
argued in the previous section that these two peaks are best explained
by the two component nature of galaxies, i.e., bulges and discs.
Fig.\ref{masstoday} shows how the stellar mass is distributed across
the $(u-r)_c-\log(n)$ plane today, and the red concentrated peak is
noticeably larger containing ($\sim 71 \pm 6$) per cent of the total
stellar mass. Spheriods (i.e., E/S0[red], c.f., Table~\ref{table2})
account for ($35 \pm 2$) per cent. Systems with both bulges and discs
must therefore account for the remaining 36 per cent. If one naively
assumes a mean $B/T \sim 0.5 \pm 0.2$, then this yields a total stellar
mass content in spheroids and bulges of ($53 \pm 7$) per cent. The discs
hence make up the remaining ($47 \pm 7$) per cent. Note that although we
truncate our distributions at $M_{\bmgc}-5 \log h=-16$ mag, the
luminosity functions are sufficiently flat to suggest a minimal
contribution to the stellar mass density and therefore dwarf systems
are negligible in terms of total stellar mass estimates (as also
inferred in \citealp{driver99}).

The very narrow distribution of the red peak (see Figs~\ref{panel2}
and~\ref{panel1}) also implies a universal continuum shape which may
only come about for old systems which formed with a universal
IMF. Mergers which lead to bulges must either occur early or without
significant star-formation i.e., dry-merging (see \citealp{graham04};
\citealp{faber} and \citealp{bell05} for fairly stringent constraints
on the level of dry-mergers at $z < 1$). Hence it seems that around
half of the stellar mass fraction ($\sim 53$ per cent) was formed
early with the remainder ($\sim 47$ per cent) forming more
recently. This at face value ties in with the well established decline
in the local star-formation rate since $z \approx 1$
(\citealp{hopkins}, \citealp{bell05}). However it is difficult to
envisage intuitively how a 2D-bimodality could occur if the whole
story consisted of a single and gradually declining star-formation
rate. One can infer an approximate age for each peak from the MGC continuum
fitting process in which the corresponding synthetic spectra ages were
15 Gyr(El), 7.4 Gyr(Sa) and 2.2 Gyr(Sc). These equate to $z>3$, $z
\approx 1$ and $z\approx 0.3$. That the Sa and Sc systems both lie in
the same blue peak points towards a common and continuous formation
process stretching and declining from at least $z \approx 1$ (i.e.,
the evolving star-forming population sought by \citealp{bell05}).

The red peak remains less certain, but because of the bimodality would
appear to share a distinct formation mechanism occurring at an earlier
time (based on the narrower red colour distribution). That the core
colour exhibits a narrower peak than the global colour suggests that
it is the cores which are coeval rather than the entire red galaxy
population. Of course red bulges can come about through early
formation of sub-units and later dry-merging (i.e., a merger event in
which minimal star-formation occurs due to the already depleted gas
reservoirs). Recent constraints on the levels of dry-merging appear
this process occurs at a low level at $z < 1$ (see \citealp{graham04};
\citealp{bell05} and also \citealp{faber}). Perhaps more persuasive is
the argument that many dry-mergers would eliminate the colour and
metallicity gradients that one sees in early-type systems (e.g.,
\citealp{labarbera03}, \citealp{depropris}). Carrying forward the
notion of a single coeval event with an implied formation peak at $z >
3$, the most obvious connection to make is with the peak in the quasar
luminosity density at $z \approx 3$ (\citealp{fan04}). The
$M_{SMBH}-L,\sigma,n$ relations (\citealp{novak}) provide additional
support for this connection (see also {\citealp{silk} for a more
theoretical basis for this connection). The key question is then
whether the bulges formed through a single monolithic collapse
(\citealp{elbs}) process, or through the rapid merger of high-mass
components in high-density environments (e.g., \citealp{menci}). An
interesting observational constraint comes from the phenomenon of
'core-depletion' in which some small fraction ($\sim 0.1$ per cent) of
the stellar population in the core region may be missing due to
ejection during SMBH coalescence. Recent constraints
(\citealp{graham04}) imply that giant spheroids ($M_R \approx -22.5
\pm 0.5$ mag), have typically undergone one dissipationless major
merger (involving the coalescence of two SMBHs). If universal, this
may point towards an early monolithic collapse (\citealp{elbs}) with
no more than one major merger. Certainly any merging scenario would
also need explain how the $M_{\mbox{SMBH}}-\sigma$
(\citealp{ferrarese}; \citealp{gebhardt}) and $M_{\mbox{SMBH}}-n$
(\citealp{graham01}) relations (see Novak et al. 2005) are
preserved. However simulations involving multiple mergers occurring
over the full lifetime do appear to be able to produce both the colour
bimodality (\citealp{menci}) and preserve the above relations
(\citealp{croton}).

Based purely on the empirical data alone, the above discussion
suggests that the two peaks are due to bulges and discs (rather than
two galaxy populations) which {\it may} be associated with the peak in
quasar activity and the broader star-formation distribution
respectively which exhibit distinct trends over distinct
time-scales. First we have the formation of the AGN/SMBH/bulge trinity,
peaking at $z \sim 3$ followed by a more extended second phase of disc
formation (presumably through splashback, infall and further
accretion). To pursue this avenue further and in a rigorous rather
than speculative manner now requires accurate bulge-disc
decompositions. These are currently in progress and results will be
reported in forthcoming papers by Allen et al. and Liske et al.

\begin{figure*}
\centering\includegraphics[width=\textwidth]{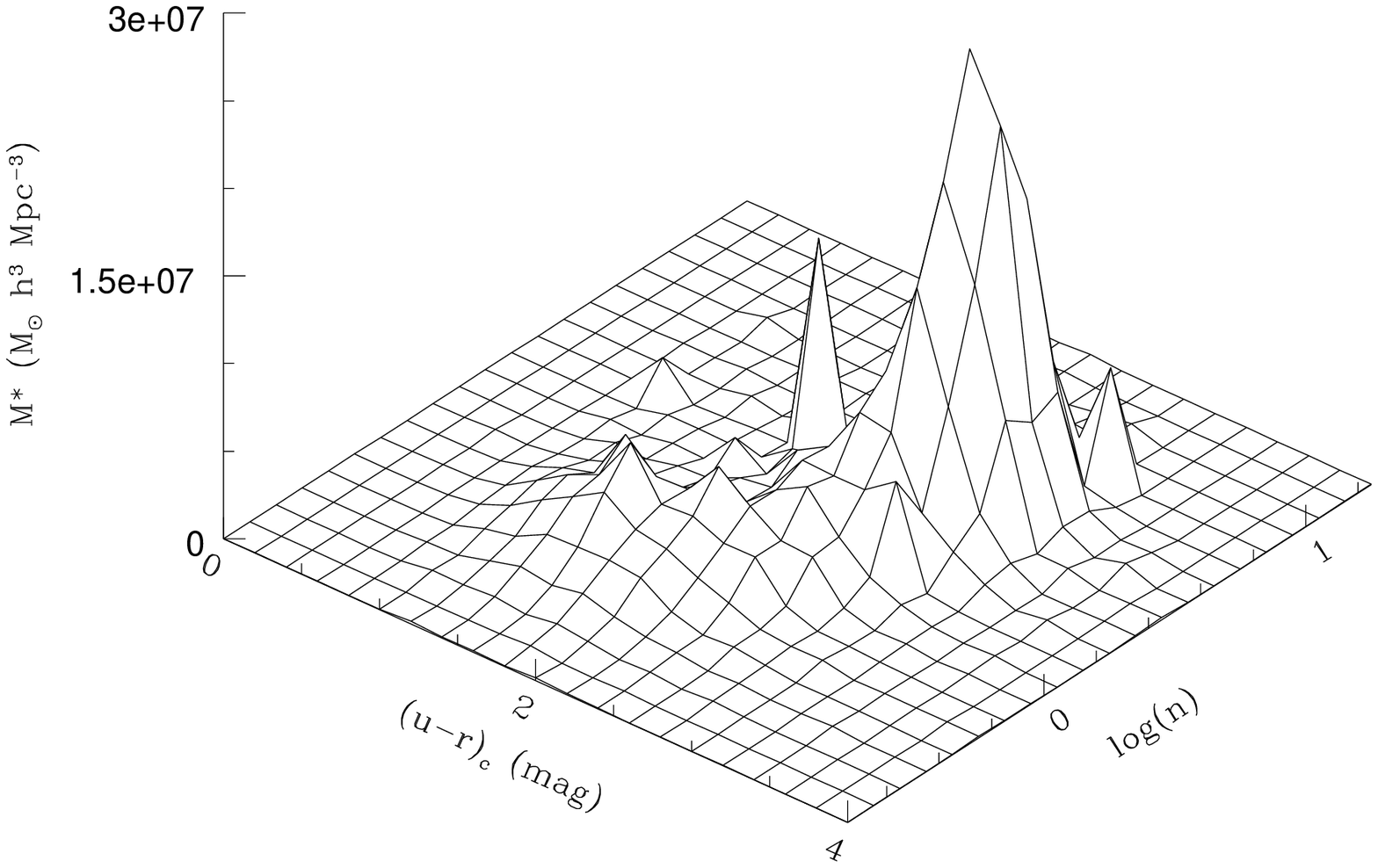}
\caption{A 3D plot showing the stellar mass density (z-axis) today
distributed across the $\log(n)-(u-r)_c$ plane.  The old red peak
clearly dominates today's stellar mass distribution with $\sim 70$ per
cent of the stellar mass contained within it.}
\label{masstoday}
\end{figure*}

\section{Summary}
We have used the Millennium Galaxy Catalogue to explore various
traditional methods for subdividing the galaxy population. The
parameters include eyeball morphological type, fitted global S\'ersic
index ($n$), 2dFGRS $\eta$-parameter, rest-$(u-r)$ colour, MGC continuum
shape, extrapolated central surface brightness, half-light radius and
stellar mass-to-light ratio. In both rest-$(u-r)$ colour and $\log(n)$
the observed and volume-corrected distributions are clearly bimodal
and the joint distribution show obvious 2D-bimodality. We find that
morphological classification is susceptible to contamination from
smooth, blue, low-luminosity systems and that overall the traditional
morphological system is wholly problematic to implement. We find that
continuum fitting of the broad-band colours leads to a good
separation of the two populations which we label ``old'' and ``young''
(and corresponds to the ``red'' and ``blue'' peaks respectively,
identified by \citealp{strateva01}).

The E/S0 systems lie predominately in the old, red peak and the Sd/Irrs
lie wholly in the young, blue peak with the Sabc class straddling both
peaks and the divide. We take this as strong evidence that {\it the galaxy
population does not consist of two classes but two components},
consistent with the classical idea of bulges and discs. The red peak
hence constitutes older bulges and the blue peak younger discs. We
advocate that this joint-bimodality (i.e., colour {\it and} concentration),
may reflect two distinct formation mechanisms occurring over different
time-scales.  We suggest that the red peak may be associated with
AGN/SMBH formation via monolithic collapse and that discs come about
through later secondary processes (splashback, infall and
accretion). Forward progress on galaxy formation demands routine
bulge-disc decomposition to enable the study of these potentially
distinctly formed components. This is currently in progress.

The Millennium Galaxy Catalogue consists of imaging
data from the Isaac Newton Telescope and spectroscopic data from the
Anglo Australian Telescope, the ANU 2.3m, the ESO New Technology
Telescope, the Telescopio Nazionale Galileo, and the Gemini
Telescope. The survey has been supported through grants from the
Particle Physics and Astronomy Research Council (UK) and the
Australian Research Council (AUS). The data and data products are
publicly available from http://www.eso.org/$\sim$jliske/mgc/ or on
request from J. Liske or S.P. Driver.

\begin{table*}
\begin{minipage}{\textwidth}
\caption{The Millennium Galaxy Catalogue $B$-band luminosity functions for
the overall sample (all) and various subdivisions (as
indicated). Errors are purely Poissonian.}
\begin{tabular}{lrccccccc} \hline 
Class & $N$ & $M^*_{\bmgc}-5\log h$ & $\phi^* (10^{-2}h^3$ & $\alpha$ &  $\chi^2$ & $j_{b_j} (10^8 L_{\odot}$ & $(\overline{g-r})$ & $\rho_M (10^{8}$\\ 
& & (mag)& Mpc$^{-3}$)   & & $(\nu =15)$ & $hMpc^{-3}$) & & M$_{\odot}$Mpc$^{-3}$\\ 
(1) & (2) & (3) & (4) & (5) & (6) & (7) & (8) & (9) \\ \hline \hline
All ($\bmgc < 19$ mag)  &3314 & $-19.60^{+0.05}_{-0.05}$ & $1.82^{+0.12}_{-0.12}$ & $-1.13^{+0.04}_{-0.04}$ & $17.1$ & 2.05 & 0.54 & $5.2\pm0.4$ \\ \hline
E/S0            &1072 & $-19.63^{+0.08}_{-0.09}$ & $0.61^{+0.06}_{-0.06}$ & $-0.92^{+0.06}_{-0.07}$ & $16.5$ & 0.62 & 0.71 & $2.7 \pm 0.4$ \\
Sabc            &1628 & $-19.43^{+0.06}_{-0.05}$ & $1.10^{+0.08}_{-0.07}$ & $-1.01^{+0.05}_{-0.05}$ & $15.0$ & 0.89 & 0.53 & $2.1\pm0.1$ \\
Sd/Irr          & 614 & $-19.49^{+0.10}_{-0.09}$ & $0.33^{+0.05}_{-0.04}$ & $-1.38^{+0.07}_{-0.06}$ & $13.4$ & 0.44 & 0.39 & $0.49\pm0.06$ \\ \hline
E/S0(red)       & 699 & $-19.47^{+0.11}_{-0.11}$ & $0.47^{+0.03}_{-0.04}$ & $-0.42^{+0.13}_{-0.12}$ & $7.3$ & 0.38 & 0.74 & $1.8 \pm 0.1$ \\
BS              & 373 & $-19.10^{+0.15}_{-0.13}$ & $0.35^{+0.07}_{-0.06}$ & $-1.10^{+0.11}_{-0.09}$ & $18.5$ & 0.24 & 0.46 & $0.9 \pm 0.4$ \\
Sd/Irr/BS       & 987 & $-19.41^{+0.09}_{-0.07}$ & $0.60^{+0.08}_{-0.06}$ & $-1.34^{+0.06}_{-0.05}$ & $15.1$ & 0.71 & 0.41 & $1.4 \pm 0.4$ \\ \hline
$\eta 1$        &1071 & $-19.40^{+0.07}_{-0.08}$ & $0.76^{+0.04}_{-0.04}$ & $-0.40^{+0.08}_{-0.09}$ & $ 9.7$ & 0.58 & 0.72 & $2.4 \pm 0.1$ \\
$\eta 2$        & 832 & $-19.19^{+0.06}_{-0.08}$ & $0.73^{+0.05}_{-0.06}$ & $-0.78^{+0.07}_{-0.08}$ & $14.5$ & 0.47 & 0.50 & $0.83 \pm 0.04$ \\
$\eta 3$        & 486 & $-19.14^{+0.13}_{-0.11}$ & $0.42^{+0.08}_{-0.06}$ & $-1.21^{+0.09}_{-0.07}$ & $ 9.1$ & 0.33 & 0.39 & $0.38 \pm 0.06$ \\
$\eta 4$        & 273 & $-19.25^{+0.17}_{-0.18}$ & $0.18^{+0.05}_{-0.04}$ & $-1.45^{+0.09}_{-0.09}$ & $14.1$ & 0.22 & 0.28 & $0.21\pm0.06$ \\
$\eta UNK$      & 652 & $-19.63^{+0.11}_{-0.11}$ & $0.32^{+0.06}_{-0.05}$ & $-1.30^{+0.08}_{-0.07}$ & $ 9.3$ & 0.44 & 0.52 & $1.4 \pm 0.4$ \\ \hline
$\eta 3+4$      & 759 & $-19.17^{+0.10}_{-0.09}$ & $0.61^{+0.09}_{-0.07}$ & $-1.29^{+0.06}_{-0.05}$ & $17.6$ & 0.54 & 0.36 & $0.6\pm0.08$ \\ \hline \hline
All ($B < 20$ mag)  &7748 & $-19.61^{+0.03}_{-0.04}$ & $1.71^{+0.05}_{-0.09}$ & $-1.18^{+0.02}_{-0.02}$ & $14.9$ & 2.03 & 0.51 & $5.2 \pm 0.4$ \\ \hline
El              &2174 & $-19.08^{+0.05}_{-0.05}$ & $0.91^{+0.02}_{-0.03}$ & $-0.13^{+0.06}_{-0.06}$ & $37.8$ & 0.55 & 0.74 & $3.3 \pm 0.4$ \\
Sa              &4366 & $-19.39^{+0.04}_{-0.04}$ & $1.23^{+0.05}_{-0.05}$ & $-1.07^{+0.02}_{-0.02}$ & $21.9$ & 1.09 & 0.47 & $1.71 \pm 0.03$ \\
Sc              &1208 & $-18.61^{+0.09}_{-0.09}$ & $0.53^{+0.08}_{-0.07}$ & $-1.47^{+0.05}_{-0.05}$ & $14.1$ & 0.37 & 0.23 & $0.24 \pm 0.06$ \\ \hline
$(u-r)_g \ge 2.1$ &3383 & $-19.32^{+0.05}_{-0.04}$ & $1.24^{+0.05}_{-0.03}$ & $-0.54^{+0.05}_{-0.03}$ & $15.8$ & 0.87 & 0.70 & $3.9 \pm 0.4$ \\
$(u-r)_g < 2.1$   &4365 & $-19.37^{+0.04}_{-0.05}$ & $1.01^{+0.05}_{-0.07}$ & $-1.35^{+0.02}_{-0.02}$ & $14.8$ & 1.17 & 0.38 & $1.31 \pm 0.07$ \\ \hline
$(u-r)_c \ge 2.35$ &2732 & $-19.15^{+0.05}_{-0.05}$ & $1.11^{+0.02}_{-0.03}$ & $-0.15^{+0.06}_{-0.05}$ & $26.4$ & 0.71 & 0.72 & $3.06 \pm 0.1$ \\
$(u-r)_c < 2.35$   &5016 & $-19.25^{+0.03}_{-0.05}$ & $1.36^{+0.06}_{-0.08}$ & $-1.28^{+0.02}_{-0.02}$ & $25.9$ & 1.29 & 0.40 & $2.2 \pm 0.4$ \\ \hline
$M^*/L_B > 2.82$ &2496 & $-19.21^{+0.05}_{-0.07}$ & $0.98^{+0.03}_{-0.05}$ & $-0.37^{+0.05}_{-0.06}$ & $28.8$ & 0.63 & 0.73 & $3.6 \pm 0.4$ \\
$M^*/L_B \le 2.82$ &5252 & $-19.48^{+0.04}_{-0.03}$ & $1.14^{+0.06}_{-0.05}$ & $-1.31^{+0.02}_{-0.02}$ & $21.0$ & 1.38 & 0.41 & $1.62 \pm 0.03$ \\ \hline
$n \ge 2.0$       &2633 & $-19.48^{+0.05}_{-0.07}$ & $0.87^{+0.03}_{-0.06}$ & $-0.66^{+0.05}_{-0.06}$ & $19.1$ & 0.75 & 0.72 & $3.35 \pm 0.4$ \\
$n < 2.0$         &5115 & $-19.35^{+0.03}_{-0.04}$ & $1.29^{+0.05}_{-0.06}$ & $-1.25^{+0.02}_{-0.02}$ & $10.2$ & 1.15 & 1.29 & $1.90\pm0.08$ \\ \hline
$\mu_o \le 19$ mag/arcsec$^{2}$  &2791 & $-19.51^{+0.06}_{-0.06}$ & $0.89^{+0.05}_{-0.05}$ & $-0.73^{+0.05}_{-0.05}$ & $18.8$ & 0.76 & 0.71 & $3.14 \pm 0.2$ \\
$\mu_o > 19$ mag/arcsec$^{2}$   &4957 & $-19.37^{+0.03}_{-0.05}$ & $1.21^{+0.05}_{-0.07}$ & $-1.26^{+0.02}_{-0.02}$ & $9.5$ & 1.25 & 0.42 & $2.1\pm0.4$ \\ \hline
\end{tabular}
\label{table2}
\end{minipage}
\end{table*}



\label{lastpage}

\end{document}